\documentclass[12pt,a4paper]{article}
\usepackage{amssymb,amsfonts,amsthm,amsmath,amscd,bbm,mathrsfs,multirow}
\usepackage{hyperref} 

\numberwithin{equation}{section}
\numberwithin{table}{section}
\def\FL{{F_{\mathrm{L}}}}
\def\phihat{{\hat{\varphi}}}
\def\varphihat{{\hat{\varphi}}}
\def\phitilde{{\tilde{\phi}}}
\def\ghat{{\hat{g}}}
\def\Bhat{{\hat{B}}}
\def\Ahat{{\hat{A}}}
\def\Chat{{\hat{C}}}
\def\RomI{{\mathrm{I}}}
\def\RomII{{\mathrm{II}}}
\def\zzbar{{z\!\!\cdot\!\!\bar{z}}}
\def\SO{{\mathrm{SO}}}
\def\SU{{\mathrm{SU}}}
\def\U{{\mathrm{U}}}

\def\Cd{{C}} 
 
\def\A{{A}} 
\def\T{{T}}
\newcommand{\gtd}[2]{g^{#1}_{\phantom{#1}#2}}

\newcommand{\Htd}[2]{H^{#1}_{\phantom{#1}#2}}

\newcommand{\Dtd}[3]{D^{#2}_{#1 #3}} 
\newcommand{\Ttd}[3]{T^{#2}_{#1#3}} 
\newcommand{\Cddt}[2]{\Cd_{#1}^{\phantom{#1}#2}}
\newcommand{\deltatd}[2]{\delta^{#1}_{\phantom{#1}#2}}
\def\realcohomK3{H^2(K3,\mathbb{R})} 
\def\Reals{\mathbb{R}}
\def\Im{\mathrm{Im}}
\def\Re{\mathrm{Re}}
\def\rmd{\mathrm{d}}
\def\dtoD{(\mathrm{d}\rightarrow D)}
\newcommand{\unit}{\ensuremath{\mathbbm{1}}}

\def\K3{{\mathrm{K3}}} \def\OK3{{O\mathrm{K3}}} 
\def\Y6{{\mathcal{Y}}}

\def\sutwo{{\mathrm{SU(2)}}}
\def\Spin{{\mathrm{Spin}}}
\def\nodd{{n_{-}}} 
\def\neven{{n_{+}}}

\def\hook{\mathbin{\rule[.2ex]{.4em}{.03em}\rule[.2ex]{.03em}{.7ex}}}

\begin{document}

\begin{titlepage}
\begin{center}

    \rightline{\small ZMP-HH/09-30}

    \vskip 1.5cm

{\Large \bf Type IIA orientifold compactification on 
SU(2)-structure manifolds}

    \vskip 1cm

    {\bf Thomas Danckaert$^{a}$ and Jan Louis$^{a,b}$}\\ 
    \vskip 1cm

    {}$^{a}${\em II. Institut f{\"u}r Theoretische Physik\\
      Universit{\"a}t Hamburg\\ 
      Luruper Chaussee 149\\ 
      D-22761 Hamburg,
      Germany}\\

    \vskip 0.4cm

    {}$^{b}${\em Zentrum f\"ur Mathematische Physik, Universit\"at
      Hamburg,\\ 
      Bundesstrasse 55, D-20146 Hamburg}

    \vskip 1cm

    {\tt thomas.danckaert@desy.de, jan.louis@desy.de} \\

  \end{center}

  \vskip 1cm

\begin{abstract}
  We investigate the effective theory of type IIA string theory on
  six-dimensional orientifold backgrounds with SU(2)-structure.  We
  focus on the case of orientifolds with $O6$-planes, for which we
  compute the bosonic effective action in the supergravity
  approximation.  For a generic SU(2)-structure background, we find
  that the low-energy effective theory is a gauged $N=2$ supergravity
  where moduli in both vector and hypermultiplets are charged.  Since
  all these supergravities descend from a corresponding $N=4$
  background, their scalar target space is always a quotient of a
  ${\SU(1,1)/\U(1) \times \SO(6,n)/\SO(6)\times \SO(n)}$ coset,
  and is therefore also very constrained.
\end{abstract}

\vfill

\end{titlepage}

\section{Introduction} 
The study of string backgrounds which are compactifications on
manifolds with $G$-structure has been of interest for some time
now. (For reviews see, for example, \cite{Grana,Fluxreviews} and
references therein.)
$G$-structure manifolds generalize Calabi-Yau spaces in that they also
admit a number of globally-defined and nowhere vanishing spinors.
These spinors, however, are no longer required to be parallel with
respect to the Levi-Civita connection.  Instead, they are parallel with
respect to a connection with torsion \cite{JoyceSalamon,Salamon2}.  As
a consequence the number of supersymmetries in the background is
unchanged but they can be spontaneously broken.
This in turn generates a potential and lifts (part of) the
vacuum degeneracy  \cite{Grana,Fluxreviews}.

So far mainly compactifications with one globally defined spinor or in
other words compactifications on SU(3)-structure manifolds were
considered \cite{Grana}. In type II theories, they lead to
four-dimensional low-energy effective theories with $N=2$
supersymmetry.  Including D-branes and orientifold planes the
supersymmetry can be further reduced to $N=1$
\cite{GrimmLouis}-\nocite{JockersLouis,MartucciSmythKoerber,Benmachiche:2006df,Acharya:2006ne,Cvetic:2007ju,Koerber:2007hd}\cite{OrientifoldReviews}.

If instead two spinors are globally defined the internal manifold has
SU(2)-structure and generically the effective theory obtained from
type II has $N=4$ supersymmetry
\cite{Waldram}-\nocite{Grana:2005sn,BovyLustTsimpis,ReidEdwardsSpanjaard,
  Triendl:2009ap}\cite{DMLST}. Including orientifold planes this
supersymmetry can be further reduced to $N=2$ or $N=1$
\cite{Grana:2005sn,Grana:2006kf,Koerber:2007hd,Andriot:2008va}.

Aspects of the low energy effective action for type II string theory
compactified on orientifolds of $\mathrm{K3}\times T^2$ have been
computed in
refs.~\cite{Dasgupta:1999ss,Tripathy:2002qw,Ferrara,Derendinger:2004jn}.
In this paper, we focus on type IIA, and calculate the bosonic $N=2$
effective action for a background manifold with SU(2)-structure and
with an $O6$ orientifold projection, within the supergravity
approximation. This is the analogue of the analysis performed in
\cite{GrimmLouis,Benmachiche:2006df} where the $N=1$ effective action
for orientifolds of SU(3)-structure compactifications was
determined. The low energy effective theory which we find is a gauged
$N=2$ supergravity where the scalar manifold ${\cal M}$ is
particularly simple and the product of the three symmetric spaces
\begin{equation}
  \label{Mresult}
  {\cal M}\ =\ \frac{\mathrm{SU}(1,1)}{\mathrm{U}(1)}
  \times \frac{\mathrm{SO}(2,n)}{\mathrm{SO}(2)\times \mathrm{SO}(n)} 
  \times \frac{\mathrm{SO}(4,m)}{\mathrm{SO}(4)\times \mathrm{SO}(m)}\ ,
\end{equation}
which descends from the scalar field space
$\mathrm{SU}(1,1)/\mathrm{U}(1)\times
\mathrm{SO}(6,n)/\mathrm{SO}(6)\times \mathrm{SO}(n)$ of $N=4$
supergravity.  The first two factors in \eqref{Mresult} are a special
K\"ahler manifold and spanned by the scalars in the vector multiplets
while the last factor is quaternionic-K\"ahler and spanned by the
scalars in the hypermultiplets.  Furthermore we find that isometries
of all three components can be simultaneously gauged when appropriate
torsion components are present.  To our knowledge, this situation has
not been encountered previously in any $N=2$ compactification of type
II string theory.\footnote{It does occur in certain heterotic
  SU(2)-structure compactification \cite{Louis:2009dq} and can
  probably also be arranged in appropriate generalizations of M-theory
  compactifications on SU(3)-structure manifolds considered in
  \cite{Micu:2006ey,Aharony:2008rx}.}

The paper is organized as follows. In section
\ref{sec:orient-type-iia-K3T2}, we consider the special case where the
internal manifold is an orientifold of the Calabi-Yau space $\K3
\times T^2$.  In order to prepare the discussion for more general
SU(2)-structure manifolds, we phrase our analysis in the formalism
introduced in \cite{Waldram,BovyLustTsimpis}.  In section
\ref{sec:type-iia-orient-general} we briefly review the orientifold
projection.  Section \ref{sec:type-iia-orient-k3-times-t2} is
concerned with finding a suitable orientifold projection on ${K3
  \times T^2}$ which preserves half of the supersymmetry.  In
\ref{sec:massless-spectrum_orientifold} we compute the massless
spectrum of the orientifolded theory and in
\ref{sec:effective-action-K3T2} determine the effective action via a
Kaluza-Klein reduction.  By performing a set of field redefinitions,
we are able to show that the scalars fields are indeed coordinates on
the scalar manifold ${\cal M}$ given in \eqref{Mresult}.  In section
\ref{sec:su2-struct-orient} we then turn to generic manifolds with
SU(2)-structure. We first describe compactifications on $G$-structure
manifolds following \cite{GLW}, and recall the properties of the
moduli space of metrics on SU(2)-structure manifolds as discussed in
\cite{Triendl:2009ap}.  We then see that most of our results from
sections \ref{sec:type-iia-orient-k3-times-t2} and
\ref{sec:massless-spectrum_orientifold} still hold in this more
general case, when we also generalize our Kaluza-Klein ansatz
appropriately.  Section \ref{sec:effective-action-su2struct} then
contains the effective action for the general case, which is a gauged
four-dimensional $N=2$ supergravity.  We then describe the obtained
gaugings in terms of the variables introduced in section
\ref{sec:effective-action-K3T2}.  The corresponding Killing
prepotentials are computed in appendix \ref{sec:calc-kill-prep}.  In
appendix \ref{sec:potential}, we show that the potential obtained from
compactification is consistent with the general formula for the
potential in $N=2$ supergravities.  Appendix
\ref{sec:spinor-conventions} contains details on the chosen spinor
conventions, appendix \ref{sec:append-cont-n=2} contains the gauge
kinetic coupling function, and we give our conclusions in section
\ref{sec:conclusions}.


\section{Orientifolds of Type IIA on K3 $\times\ T^2$}
\label{sec:orient-type-iia-K3T2}

\subsection{Type IIA Orientifolds}
\label{sec:type-iia-orient-general}
Let us start by recalling the orientifold projection for type IIA
theories \cite{OrientifoldReviews}.  {}From a
world-sheet 
perspective, orientifolds arise by modding out the string theory
by a discrete involutive symmetry $S$. This symmetry includes the
map $\Omega_p$ which inverts orientation of the string world sheet
(parametrized by $\sigma$ and $\tau$) according to
\begin{equation}
 \label{eq:Omega}
\Omega_p:\ (\sigma,\tau) \rightarrow
 (2\pi-\sigma,\tau)\ .
\end{equation} 
$\Omega_p$ is such that it exchanges left- and right-moving string
modes.  For the fermionic modes in type IIA, this means spinors of
opposite target-space chirality must be mapped to each other.  In
order to do this consistently, one has to combine $\Omega_p$ with an
involution $\sigma$ of the target space, which inverts target space
orientation \cite{OrientifoldReviews}.\footnote{To be consistent with
  the standard notation we use $\sigma$ to also denote this involution
  which, however, has nothing to do with the world-sheet coordinate
  $\sigma$.} Locally, such a $\sigma$ can be thought of as an odd
number of reflections along tangent space directions.  Depending on
the number of reflections, the transformation properties of the
fermionic modes change and one may have to add an extra operator
$(-1)^\FL$ in order to ensure $S^2=1$ for all states
\cite{OrientifoldReviews}.  More straightforwardly, the number
of flipped directions also determines the generic dimension of the
orientifold planes, since these lie at the fixed-point loci of
$\sigma$.  Thus, one arrives at the following possibilities for
$Op$-planes and the corresponding projections:
\begin{equation}
 \begin{tabular}[h!]{ll} O2,O6: & $\quad S=(-1)^\FL\Omega_p\sigma$\ ,\\ 
O0,
 O4, O8: & $\quad S=\Omega_p\sigma$\ .
 \end{tabular}
\end{equation}
(More details of this projection are given in appendix
\ref{sec:spinor-transf-prop}.)

In type IIA string theory, the massless ten-dimensional bosonic
spectrum consists of the metric $\hat{g}$, the dilaton
$\hat{\varphi}$, and the 2-form $\hat{B}$, all in the NS-sector, and a
one- and three-form field $\Ahat$ and $\Chat$ in the RR-sector.  The
orientifold map $S$ acts on these fields by the pull-back $\sigma^*$
of the target-space involution $\sigma$, combined with extra minus
signs, which can be deduced from the world-sheet description of each
field. Altogether one has
\begin{equation}
  \label{eq:transformation_massless_fields_Omegasigma}
  \Omega_p\sigma:\left\{
    \begin{array}{lcr} \hat{\varphi} &\rightarrow
      &\sigma^*(\hat{\varphi})\\ 
      \hat{g} &\rightarrow &\sigma^*(\hat{g})\\
      \Bhat &\rightarrow &-\sigma^*(\Bhat)\\ 
      \Ahat &\rightarrow
      &\sigma^*(\Ahat)\\ 
      \hat{C} &\rightarrow &-\sigma^*(\hat{C})
    \end{array}\right., \qquad (-1)^\FL:\left\{
    \begin{array}{lcr} \phihat & \rightarrow & \phihat\\ 
      \ghat &
      \rightarrow & \ghat\\ 
      \Bhat & \rightarrow & \Bhat\\ 
      \Ahat
      &\rightarrow & -\Ahat\\ 
      \Chat &\rightarrow & -\Chat
    \end{array}\right..
\end{equation} 
$(-1)^\FL$ only acts on the R-R fields, since they are built from the
tensor product of one left- and one right-moving world-sheet spinor.
The transformation properties under $\Omega_p\sigma$ can be derived by
writing the NS-NS modes as symmetric or antisymmetric products of
left-and right-moving bosonic oscillators, and the R-R modes as spinor
bilinears, and switching left- and right-moving
modes \cite{OrientifoldReviews}. 

\subsection{Orientifold action on $\K3 \times T^2$}
\label{sec:type-iia-orient-k3-times-t2}
Let us now consider backgrounds which include orientifolds of the
six-dimensional Calabi-Yau manifolds $\K3 \times T^2$ and first
determine the appropriate orientifold projection for this case.  K3 is
the unique four-dimensional manifold with SU(2)-holonomy and thus
$\K3\ \times\ T^2$ can be viewed as a special case of a
six-dimensional manifold with SU(2) structure.  We will see that it
already exhibits many features of a generic SU(2)-structure manifold
that we will analyze in section \ref{sec:su2-struct-orient}.

A Ricci-flat metric on $\K3\ \times\ T^2$ admits 2 covariantly constant
spinors $\eta^i, i=1,2$. As a consequence the two parameters
$\varepsilon^{\RomI, \RomII}$ of the 
unbroken ten-dimensional type II supersymmetry transformations may be
decomposed as
\begin{equation}
  \label{eq:susy-param-decomposition}
  \begin{aligned}
    \varepsilon^\RomI_{10} &= \varepsilon^I_{i+}\otimes \eta^i_+ + \varepsilon^I_{i-}\otimes \eta^i_-\, ,\\
    \varepsilon^\RomII_{10} &= \varepsilon^\RomII_{i+}\otimes \eta^i_- -\varepsilon^\RomII_{i-}\otimes\eta^i_+\, ,
  \end{aligned}\qquad i =1,2\ ,
\end{equation}
where the $\varepsilon_i$ and the $\eta^i$ are $\Spin(1,3)$, resp.\
$\Spin(6)$ Weyl spinors.  The minus sign in the second line of
\eqref{eq:susy-param-decomposition} is due to our choice for the
Majorana condition and the fact that $\varepsilon^\RomII_{10}$ has
chirality -1 (see appendix \ref{sec:spinor-conventions} for details on
the chosen conventions).  We see that, from a four-dimensional
perspective, the unbroken supersymmetries feature the four parameters
$\varepsilon^{\RomI}_i, \varepsilon^\RomII_i$.  Therefore type IIA
string theory in a K3 $\times\ T^2$ background is described at low
energies by a four-dimensional $N=4$ supergravity theory
\cite{Duffetal,ReidEdwardsSpanjaard}.

In this section we aim at constructing an $N=2$ theory by including an
appropriate orientifold projection in this setup.  Preserving $N=2$
supersymmetry requires that $\sigma$ leaves half of the
supersymmetries invariant.  For concreteness we consider 
a background with $O6$-planes, for which  $\sigma$ acts on the six-dimensional
spinors $\eta^i$ as follows \cite{Benmachiche:2006df,Koerber:2007hd}
\begin{equation}
  \label{eq:involution_spinors}
  \sigma^*(\eta^i_\pm) =\pm\, \eta^i_\mp\ .
\end{equation} 
(We derive the form of this projection in our conventions in appendix~\ref{sec:spinor-conventions}.)
In general, one can add a multiplication by a phase $e^{i\theta}$ to
the action of $\sigma$, which has the effect of rotating the
$O$-planes.  However, in the case of a single $O$-plane, one can
always choose suitable variables in which the phases disappear.

The K\"ahler form $J$ and the holomorphic 2-form $\Omega$ on K3 can
be expressed in terms of the globally defined spinors $\eta^i$ as 
\cite{BovyLustTsimpis,Waldram}\footnote{Let us
summarize our conventions for the different coordinates: ten-dimensional
coordinates are labeled $X^M, M=0,...,9$ while
four-dimensional space-time coordinates are labeled by $x^\mu, \mu=0,...,3$.
The real coordinates on the internal manifold (here $\K3\times
T^2$) are labeled $Y^m$  which are split into 
the coordinates $y^i,
i=1,2$ on $T^2$  and the (real) coordinates $z^a,
a=1,...,4$ on K3.\label{fn:coords} }
\begin{equation}
  \label{eq:su2struct_2forms_K3xT2}
  \begin{aligned}
    & J := \tfrac{i}{4}{
      (\eta^{1\dag}_{-}\gamma_{mn}\eta^1_{-} 
      -\eta^{2\dag}_-\gamma_{mn}\eta^2_{-})\,
      \rmd Y^m \wedge \rmd Y^n}
    &= \tfrac{1}{2}J_{ab}\rmd z^a \wedge \rmd z^b\, ,\\
    &\Omega := \tfrac{i}{2}\eta^{1\dag}_-\gamma_{mn}\eta^2_-\, \rmd Y^m \wedge \rmd Y^n
    &=\tfrac{1}{2}\Omega_{ab} \rmd z^a\wedge \rmd z^b\, .
  \end{aligned}
\end{equation}
Using \eqref{eq:involution_spinors} one can infer the following transformation
properties \cite{OrientifoldReviews}:
\begin{equation}
  \label{eq:sigma_compl_struct_hol_2form}
  \begin{aligned}
    \sigma^*(J) &= -J,\\ 
    \sigma^*(\Omega)&= -\bar{\Omega}.
  \end{aligned}
\end{equation}
In addition, the basis one-forms on the torus can also be expressed
as a bispinor in the $\eta^i$ via \cite{BovyLustTsimpis}
\begin{equation}
  \label{Kdef}
  K := \eta^{2\dag}_- \gamma_m \eta^1_+\, \rmd Y^m = \rmd y^2 + i\rmd y^1\ ,
\end{equation}
 where $\gamma_1, \gamma_2$ are the gamma-matrices in
the torus directions.  Here, the transformations
\eqref{eq:involution_spinors} act as 
\begin{equation}
  \label{eq:orientifold_involution_torus}
  \sigma^*(K) = \bar{K}\ ,
\end{equation}
which implies $\sigma^*(\rmd y^1) = - \rmd y^1$ and $\sigma^*(\rmd
y^2) = \rmd y^2$.

\subsection{Massless spectrum} 
\label{sec:massless-spectrum_orientifold}
Let us now determine the massless spectrum of the orientifolded theory
and assign it to $N=2$ multiplets. As in any Calabi-Yau
compactification, the massless modes of the four-dimensional theory
are obtained by expanding the ten-dimensional fields in harmonic modes
on K3 $\times\ T^2$.  On K3 there are the constant function, the 22
harmonic 2-forms $\omega^\alpha(y)$ and one harmonic 4-form.  On
$T^2$, the ``harmonic modes'' are just the constant functions and
forms.  Therefore the ten-dimensional (hatted) fields can be expanded
as
\begin{equation}
  \label{eq:KK_expansion_formfields}
  \begin{aligned}
    \Ahat =& \ \A + \A_i\nu^i\ , \quad i=1,2
    \\ 
    \Bhat =&\ B +
    B_{ i}\wedge\nu^i + \tfrac{1}{2} B_{ij}\, \nu^i\wedge \nu^j + B_\alpha
    \omega^\alpha(y)\ ,\quad \alpha=1,...,22\ ,
    \\ 
    \Chat =&\ \Cd + (\Cd_i
    -\A\wedge B_i)\wedge \nu^i + \tfrac{1}{2}(\Cd_{ij}-\A\, B_{ij})\wedge
    \nu^i\wedge\nu^j \\ & +(\Cd_{\alpha}-\A\, B_\alpha)\wedge
    \omega^\alpha + \Cd_{i\alpha}\nu^i\wedge
    \omega^\alpha\ ,
  \end{aligned} 
\end{equation}
where the ``vielbein'' one-forms $\nu^i$ are defined as $\nu^i=\rmd
y^i -\gtd{i}{\mu}\rmd x^\mu$ with $\gtd{i}{\mu}$ being the appropriate
off-diagonal metric component. All other (unhatted) variables denote
four-dimensional fields.  In this basis the metric is block-diagonal
and given by
\begin{equation}
  \label{eq:10D_metric_decomposition}
  \hat{\rmd s}^2 = g_{\mu\nu}\rmd
  x^\mu \rmd x^\nu + g_{ij}(\rmd y^i -\gtd{i}{\mu}\rmd x^\mu)
  (\rmd y^j -\gtd{j}{\nu}\rmd x^\nu) + g_{ab}\rmd z^a \rmd z^b\ ,
\end{equation}
where $g_{\mu\nu}$, $\gtd{i}{\mu}$ and $g_{ij}$ only depend on $x^\mu$
while $g_{ab}$ is the metric on K3 which also depends on the K3
coordinates $z^a$.

In order to determine the spectrum of the orientifolded theory, we
have to project the spectrum onto the modes which are even under the
map $S=(-1)^\FL\Omega \sigma$.  The pull-back of $\sigma$ splits the
modes in the Kaluza-Klein expansion
\eqref{eq:KK_expansion_formfields} into an
eigenspace with eigenvalue $+1$, and an eigenspace with eigenvalue
$-1$, which we will refer to as the even and odd eigenspaces.  The
forms $\rmd x^\mu$ are even, since $\sigma$ does not act on the
non-compact directions. For $T^2$ we determined below
\eqref{eq:orientifold_involution_torus} that $\rmd y^1$ is odd and
$\rmd y^2$ is even.  The harmonic 2-forms $\omega^\alpha$ on K3 split
into an odd eigenspace $H^{2,-}$ of dimension $\nodd$ and an even
eigenspace $H^{2,+}$, of dimension $\neven$.\footnote{The numbers
  $\neven$ and $\nodd$ depend on the involution $\sigma$, which we do
not specify here.}
We can always choose a basis of forms in
$H^2$ which consists of eigenvectors of the involution
$\sigma$. This basis then splits
into a basis of $H^{2,+}$ and a basis of $H^{2,-}$, which we label as
follows:
\begin{equation}
  \begin{aligned}
    \label{Hdef}
    H^{2,+} &= \mathrm{span}\left\{\omega^A\right\}\ ,\qquad
    A=1,...,\neven\ ,\\ 
    H^{2,-} &=
    \mathrm{span}\left\{\omega^P\right\}\ ,\qquad  P=1,...,\nodd\ ,
  \end{aligned}
\end{equation} 
with $\neven+\nodd=22$.  Since the wedge product
and the pull-back $\sigma^*$ commute, it is easy to determine the
parity with respect to $\sigma^*$ of products of the $\omega^\alpha$
and $\rmd y^i$.

Using \eqref{eq:transformation_massless_fields_Omegasigma} together
with the action of $\sigma^*$ that we have just determined we can now
determine the orientifold spectrum by projecting onto those modes
which are invariant under $S=(-1)^\FL\Omega_p\sigma$.  Let us start
with the components of the metric $\ghat$ on $\Reals^{1,3}\times T^2$.
It is slightly more transparent to do the projection for the metric
components in the coordinate basis which uses $\rmd x^\mu, \rmd y^i$
as differentials rather than the ``vielbein'' basis $\rmd x^\mu,
\nu^i$ of \eqref{eq:10D_metric_decomposition}.  Let us define
\begin{equation}
  \label{eq:metric-components-compare}
  \ghat_{\mu\nu}=g_{\mu\nu} +
  g_{ij}\gtd{i}{\mu}\gtd{j}{\nu}\ ,\qquad
  \ghat_{i\mu}=-g_{ij}\gtd{i}{\mu}\ ,\qquad \ghat_{ij}=g_{ij}\ ,
\end{equation}
such that $\hat{\rmd s}^2 = \ghat_{\mu\nu}\rmd x^\mu \rmd x^\nu +
\ghat_{ij}\rmd y^i \rmd y^j+2\ghat_{i\mu}\rmd y^i \rmd x^\mu+
g_{ab}\rmd z^a \rmd z^b$.

$S$ maps the metric $\ghat$ to $\sigma^*(\ghat)$, so we have to
project out the modes with odd parity under the action of $\sigma^*$.
Since we are restricting ourselves to the metric on
$\Reals^{1,3}\times T^2$ for now, this means that we only keep the
even forms $\rmd x^\mu \rmd x^\nu, \rmd x^\mu \rmd y^2,
\rmd y^1\rmd y^1$ and $\rmd y^2\rmd y^2$ or in other
words the components $\ghat_{\mu\nu}, \ghat_{2\mu},\ghat_{11}$ and
$\ghat_{22}$ remain in the spectrum.  Using
\eqref{eq:metric-components-compare} to return to the ``vielbein''
frame, we see that we are left with the components $g_{\mu\nu},
g_{11}$ and $g_{22}$.  The $\gtd{i}{\mu}$, which are related to the
metric $\ghat$ by $\gtd{i}{\mu}= g^{ij}\ghat_{j\mu}$ are reduced
to $\gtd{1}{\mu}=0$, $\gtd{2}{\mu}=g^{22}\ghat_{2\mu}$ by the
orientifold projection.  Now it is easy to see that $\nu^1=\rmd y^1$
is odd while $\nu^2= \rmd y^2 -\gtd{2}\mu\rmd x^\mu$ is even.

Let us continue to impose the orientifold projection on the other
fields in the spectrum.
\eqref{eq:transformation_massless_fields_Omegasigma} implies that the
two-form field $\Bhat$ has to transform as $\Bhat\rightarrow
-\sigma^*(\Bhat)$, so that only odd modes survive in the expansion of
$\hat B$ in 
\eqref{eq:KK_expansion_formfields}.  These are the coefficients of
$\nu^1\wedge \rmd x^\mu, \nu^1\wedge\nu^2$ and $\omega^P$ or in other
words the components $B_{1\mu}, B_{12}$ and $B_P$.  $\Ahat$
similarly transforms as $\Ahat\rightarrow-\sigma^*(\Ahat)$ so that
again only the $\sigma^*$-odd component $A_1$ survives.  The
three-form $\Chat$ transforms as $\Chat \rightarrow \sigma^*(\Chat)$
which implies that the even modes $\Cd_{2\mu\nu}, \Cd_{A\mu},
\Cd_{1P}, \Cd_{2A}$ and $\Cd_{\mu\nu\rho}$ remain in the spectrum.
($\Cd_{\mu\nu\rho}$, however, contains no dynamical degrees of
freedom.)  Finally from the dilaton only the even $x$-dependent scalar
field $\varphi$ is kept.

This concludes the truncation of the modes coming from
\eqref{eq:KK_expansion_formfields}.  The reduction of the action for
the K3 metric $g_{ab}$ is slightly more complicated.  A Ricci-flat
metric on K3 is determined, up to the global volume factor, by its
hyperk\"ahler structure.  The hyperk\"ahler structure in turn is
determined by the subspace $\Sigma$ of the second cohomology class
$\realcohomK3$ spanned by the two-forms $J$, $\mathrm{Re}\Omega$ and
$\mathrm{Im}\Omega$ defined in \eqref{eq:su2struct_2forms_K3xT2}, or
equivalently, the space of self-dual harmonic two-forms on K3
\cite{Aspinwall:1996mn}.  The second cohomology class of K3 is a
22-dimensional vector space, equipped with a metric of signature
(3,19) via the intersection product.  One can express the intersection
product in terms of the basis $\left\{\omega^\alpha\right\}$ of
harmonic two-forms:
\begin{equation}
  \label{eq:intersectionproduct_K3}
  \eta^{\alpha\beta} =
  \int_{\K3}\omega^\alpha\wedge\omega^\beta\ ,\qquad \alpha,
  \beta=1,\ldots,22\ .
\end{equation}
The space of self-dual forms $\Sigma$ is then a 3-dimensional subspace
of the second cohomology class, spanned by forms with a positive
self-intersection number, which one can regard as a subspace of
$\Reals^{3,19}$ spanned by vectors with positive norm.\footnote{In the
  remainder of the article we call these subspaces ``spacelike
  subspaces''.}  As a consequence the moduli space of Ricci-flat
metrics on K3 is the Grassmannian
\begin{equation}
  \label{eq:K3metricmoduli}
  \mathcal{M}_\K3 = \frac{\mathrm{SO}(3,19)}
  {\mathrm{SO}(3)\times \mathrm{SO}(19)} \times
  \Reals^+ \ ,
\end{equation}
up to a quotient by the discrete group of isomorphisms on K3
\cite{Aspinwall:1996mn}.  The factor $\Reals^+$ is the volume of the
K3 surface, which we denote as
\begin{equation}
  \label{eq:K3volume}
  e^{-\rho} = \int_{\K3}\sqrt{\mathrm{det}(g_{ab})} \ .
\end{equation}
The remaining moduli can be conveniently encoded in a matrix
$\Htd{\alpha}{\beta}$, which determines the action of the Hodge
$*$-operator on the harmonic 2-forms via
\begin{equation}
  \label{eq:K3_metric_moduli_matrix}
  *\omega^\alpha = \Htd{\alpha}{\beta}\omega^\beta\ .
\end{equation}
We can also describe the matrix $\Htd{\alpha}{\beta}$ in terms of
three orthonormal (with respect to $\eta^{\alpha\beta}$) vectors
$\xi^{x}_{\phantom{x}\alpha},x=1,2,3$, which parametrize the
variations of the two-forms $J$ and $\Omega$ \cite{Louis:2009dq}
\begin{equation}
  \label{eq:xi_2forms}
  \begin{aligned}
    J &= \sqrt{2}\, e^{-\frac{\rho}{2}}\xi^1_{\phantom{1}\alpha}\omega^\alpha \, ,\\
    \Omega &=
    \sqrt{2}\, e^{-\frac{\rho}{2}}(\xi^2_{\phantom{2}\alpha}\omega^\alpha
    + i\xi^3_{\phantom{3}\alpha}\omega^\alpha)\, ,
  \end{aligned}
\end{equation}
Since $J$ and $\Omega$ span the subspace $\Sigma$ of self-dual
harmonic two-forms, $\Htd{\alpha}{\beta}$ takes the form
\cite{Louis:2009dq}
\begin{equation}
  \label{eq:moduli_matrix_xi}
\Htd{\alpha}{\beta} =
  -\deltatd{\alpha}{\beta} +2\xi^{x\alpha}\xi^x_{\phantom{x}\beta} \ ,
\end{equation}
where $\xi^{x\alpha}= \eta^{\alpha\beta}\xi^x_{\phantom{x}\beta}$.  

The next step is to determine the $S$-invariant subspace of
$\mathcal{M}_\K3$ or in other words determine the $S$-invariant
deformations of the K3-metric.  From the transformation properties
given in \eqref{eq:transformation_massless_fields_Omegasigma} we learn
that these are the deformation which are invariant under the action of
$\sigma^*$.  Since $\sigma$ is an isometry, it leaves the Hodge
$*$-operator invariant.  It follows that the Hodge $*$-operator only
acts within each of the eigenspaces $H^{2,\pm}$ defined in
\eqref{Hdef}.  From \eqref{eq:K3_metric_moduli_matrix} we then
immediately conclude that the matrix $\Htd{\alpha}{\beta}$ has to be
block-diagonal, i.e.
\begin{equation}
  \label{eq:hodge_star_splitting}
  \Htd{\alpha}{\beta} = \left(
    \begin{array}{cc}
      \Htd{A}{B} &0\\ 0 &\Htd{P}{Q}
    \end{array}\right)\ ,
\end{equation}
where $A,B=1,...,\neven$ and $P,Q=1,...,\nodd$ label the even and odd
two-forms, respectively.  The intersection product
\eqref{eq:intersectionproduct_K3} is a topological invariant, so it
remains unchanged under the action of any diffeomorphism, and thus,
more specifically, under the involution $\sigma$.  Therefore, also
$\eta^{\alpha\beta}$ has the block-diagonal form
\begin{equation}
  \label{eq:intersectionform_split}
  \eta^{\alpha\beta}=\left(\begin{array}{cc}
      \eta^{AB} & 0\\ 0 & \eta^{PQ}
    \end{array}\right)\ .
\end{equation}
As we already
recalled $J, \Re \Omega$ and $\Im \Omega$ span $\Sigma$ and thus have
positive self-intersection number.  It follows from the transformation
properties \eqref{eq:sigma_compl_struct_hol_2form} that $\Im \Omega$
lies in $H^{2,+}$, whereas $J$ and $\Re \Omega$ lie in $H^{2,-}$.
Together, these facts imply that the intersection form $\eta^{AB}$
on $H^{2,+}$ has signature $(1,\neven-1)$, whereas $\eta^{PQ}$ has
signature $(2,\nodd-2)$.

The reduction of $\Htd{\alpha}{\beta}$ to a block-diagonal form
corresponds to the following reduction of the parameter space of the
$\xi^x_{\phantom{x}\alpha}$: the choice of three orthonormal vectors
$\xi^x \in \Reals^{3,19}$ is reduced to a choice of one unit
vector $\xi^3_{\phantom{1}A} \in \Reals^{1,\neven-1}$ (with $\neven -1$
degrees of freedom) and two orthogonal unit vectors
$\xi^{1}_{\phantom{1}P},\xi^{2}_{\phantom{2}P} \in
\Reals^{2,\neven-2}$ (with $2(\neven-2)$ degrees of freedom). In other
words, the matrices $\Htd{A}{B}, \Htd{P}{Q}$ are given by
\begin{equation}
  \label{eq:reduced_moduli_matrix_xi}
  \Htd{A}{B}=-\deltatd{A}{B}
  +2\xi^{3A}\xi^3_{\phantom{3}A}, \qquad
  \Htd{P}{Q}=-\deltatd{P}{Q}
  +2(\xi^{1P}\xi^1_{\phantom{1}Q}+\xi^{2P}\xi^2_{\phantom{2}Q}).
\end{equation}
We see that, for the metric to be invariant under the orientifold
projection, the spacelike three-plane $\Sigma$ must be a product of a
one-dimensional spacelike subspace in $H^{2+}$ and a two-dimensional
spacelike subspace in $H^{2,-}$.  This means that the Grassmannian in
equation \eqref{eq:K3metricmoduli} is reduced to the product of two
Grassmannians.  Together with the volume factor, this accounts for the
moduli space
\begin{equation}
  \label{eq:K3_metric_moduli_space_orientifolded}
\mathcal{M}_{\OK3}\ =\ \frac{
    \mathrm{SO}(1,\neven-1)}{\mathrm{SO}(\neven-1)} \times\frac{\mathrm{SO}(2,\nodd-2)}{\mathrm{SO}(2)\times
    \mathrm{SO}(\nodd-2)} \times \Reals^+\ .
\end{equation} 

\begin{table}
  \centering
  \begin{tabular}{|c||c|c|c|}
    \hline & $j=2$&$j=1$&$j=0$\\ 
    \hline
    \multirow{4}{*}{$\ghat$}&$g_{\mu\nu}$ & $\gtd{2}{\mu} $ & $g_{22},
    g_{11}$\\ 
    && & $\Htd{A}{B}$\\ 
    && & $\Htd{P}{Q}$\\ 
    && & $\rho$\\
    \hline $\phihat$&& & $\varphi$\\ 
    \hline
    \multirow{2}{*}{$\Bhat$}&&$B_{\mu1}$& $B_{12}$\\ 
    &&&$B_P$\\ 
    \hline
    $\Ahat$&&&$\A_1$\\ 
    \hline \multirow{3}{*}{$\Chat$}&&$\Cd_{\mu
      A}$&$\Cd_{2A}$\\ 
    &&&$\Cd_{1P}$\\ 
    &&&$\Cd_{2\mu\nu}$\\ 
    \hline
  \end{tabular}
  \caption{This table lists the massless fields which survive the 
$O6$ orientifold projection.  The hatted fields are the massless
ten-dimensional fields while the unhatted fields are the 
massless modes in four space-time dimensions
with $j$ indicating their (four-dimensional)
    spin. The indices
  $A,B=1,...,\neven$ label components from the expansion in even
  two-forms, the indices $P,Q=1,...,\nodd$ correspond to odd
  two-forms.}\label{tab:K3_orienti_spectrum}
\end{table} 

To summarize, 
we determined the massless bosonic modes which survive 
the orientifold projection and assembled them in 
Table~\ref{tab:K3_orienti_spectrum}.  As we will discuss in more detail in
the coming sections, these fields match the bosonic content of a
four-dimensional $N=2$ supergravity theory which contains, apart from the
gravity multiplet, $\neven +1$ vector multiplets and $\nodd$
hypermultiplets. However, we can already 
anticipate the field content of these multiplets:
\begin{itemize}
\item The gravity multiplet contains the metric $g_{\mu\nu}$ and the
graviphoton $\gtd{2}{\mu}$.
\item  The $\neven$ vector fields $\Cd^A$, the $\neven$ real scalars 
$\Cd_2^{\phantom{2}A}$, the $(\neven-1)$ degrees of freedom from
$\xi^{3A}$, and $e^{-2\varphihat-\rho}g_{22}$ together form $\neven$ vector multiplets.
\item The vector field $B_1$, the product $g_{11}g_{22}=e^{-2\eta}$
and $B_{12}$ form one more vector multiplet.
\item  The $\nodd$ scalars
$B^P$, the $\nodd$ scalars $\Cd_{1P}$ and the $2(\nodd-2)$ degrees of freedom contained in
$\Htd{P}{Q}$ assemble in $(\nodd-1)$ hypermultiplets.
\item An additional hypermultiplet arises as the Poincar\'e dual of the
tensor multiplet containing the scalars $\A_1$ and $e^{-2\varphihat-\rho}g_{11}$, the K3 volume factor $\rho$, and the two-form $\Cd_{2}$.
\end{itemize}
We note that the dilaton $\varphihat$ is a combination of scalars from
the vector and hypermultiplets.  This implies that both sectors
receive string loop corrections, as is the case in type I
compactifications \cite{Antoniadis:1997nz}.

\subsection{Effective action}\label{sec:effective-action-K3T2}
We can now compute the effective four-dimensional action for the
orientifolded theory.  The starting point is the bosonic action of
ten-dimensional type IIA supergravity given by
\cite{Polchinski:1998rr}
\begin{equation}\begin{aligned}
 \label{eq:typeIIA_eff_action}
 S_{IIA}= & \int e^{-2\phihat}\left(
   \rmd^{10}x \sqrt{-\ghat}(\hat{R} 
   +4\partial_M\phihat \partial^M\phihat) 
   +\tfrac{1}{2}\rmd \Bhat 
   \wedge *\rmd \Bhat \right) \\ 
 &+\tfrac{1}{2}\int 
 \left( 
   \rmd \Ahat
   \wedge * \rmd \Ahat + \tilde{F}_4 \wedge * \tilde{F}_4
 \right) 
 +\tfrac{1}{2}\int 
 \Bhat \wedge \rmd \Chat \wedge \rmd \Chat\ ,
\end{aligned}\end{equation} 
with the field strength
\begin{equation}
 \label{eq:Ftilde}
\tilde{F}_4 = \rmd \Chat - \Ahat\wedge \rmd \Bhat\ .
\end{equation}
Substituting the Kaluza-Klein expansion of
\eqref{eq:KK_expansion_formfields} together with the orientifold
projection as determined in the previous section into the action
\eqref{eq:typeIIA_eff_action} and integrating over $\K3 \times T^2$,
we obtain\footnote{We do not give the computation here but refer the
  reader to ref.~\cite{Louis:2009dq} for further details of the
  reduction in the NS-sector and to
  ref.~\cite{ReidEdwardsSpanjaard,DMLST} for the reduction in the
  RR-sector. The six-dimensional action obtained from K3 compactifications of type IIA is given in \cite{Duffetal}, while background fluxes are turned on in \cite{Haack:2001iz}.}
\begin{equation}
  \label{eq:K3xT2_orienti_action}
  \begin{aligned}
  S_{\mathrm{kin}}=\quad &\int \rmd^4 x \sqrt{-g}
  \left(
    \tfrac{1}{2}R 
    -\partial_\mu(\varphihat +\tfrac12 \rho +\tfrac{1}{2}\eta)
    \partial^\mu(\varphihat +\tfrac12 \rho +\tfrac{1}{2}\eta)\right.\\
  &\qquad\left.
    +\tfrac{1}{8}(\partial_\mu e^{\rho}\partial^\mu e^{-\rho}
    +\partial_\mu g_{11}\partial^\mu g^{11} +\partial_\mu
    g_{22}\partial^\mu g^{22})
  \right.\\
  &\qquad\left.
    +\tfrac{1}{16}(\partial_\mu \Htd{A}{B}\partial^\mu\Htd{B}{A}
    +\partial_\mu \Htd{P}{Q}\partial^\mu\Htd{Q}{P})\right)\\
  +&\tfrac{1}{2}\int \Big(
    \tfrac12{e^\rho} H^{PQ}(\rmd B_P \wedge * \rmd B_Q)
    +\tfrac12{e^{2\eta}}\rmd B_{12}\wedge*\rmd B_{12} \\
    &\qquad +e^{-2\varphihat -\rho -\eta}g_{22}\rmd g^2 \wedge * \rmd g^2 \\
    &\qquad +e^{-2\varphihat -\rho -\eta}g^{11}(\rmd B_1 
      + \rmd g^2 B_{12})\wedge * (\rmd B_1 + \rmd g^2B_{12}) 
  \Big)\\
  +\tfrac{1}{2}&\int \Big(
    \tfrac12{e^{2\varphihat}}g^{11}\rmd \A_1 \wedge * \rmd \A_1 
    +\tfrac12{e^{2\varphihat+\rho}}H^{fAB}g^{22}
    \rmd \Cd_{2A}\wedge * \rmd \Cd_{2B}\\ 
    &\qquad +\tfrac12{e^{2\varphihat+\rho}} H^{PQ}g^{11}
      (\rmd \Cd_{1P}+\A_1\rmd B_P)
      \wedge *(\rmd \Cd_{1Q} +\A_1\rmd B_Q) \\
    &\qquad
    + e^{-\eta} H^{AB} (\rmd \Cd_A - \rmd g^2\Cd_{2A}) 
    \wedge *(\rmd \Cd_B - \rmd g^2\Cd_{2B}) \\ 
    &\qquad
    +2e^{-2\varphihat-2\eta-2\rho}g^{22}\rmd\Cd_2 \wedge * \rmd \Cd_2 \\
    &\qquad
    +4e^{-4\varphihat-3\eta-3\rho}(\rmd \Cd -\rmd g^2\wedge \Cd_2)\wedge *
  (\rmd \Cd -\rmd g^2\wedge \Cd_2) \Big)\\
  +\tfrac{1}{2}&\int \Big(
    -\eta^{AB}B_{12}\rmd \Cd_A\wedge \rmd \Cd_B 
    -2 \eta^{PQ}\rmd \Cd_2 \wedge B_P\rmd\Cd_{1Q}\\ 
    &\qquad + \eta^{AB} (\rmd B_1 + \rmd g^2B_{12})
      \wedge (2 \rmd \Cd_A - \rmd g^2\Cd_{2A})\Cd_{2B}
    \Big)\ .
  \end{aligned}
\end{equation}
$\eta$ encodes the volume of the torus $T^2$ and is defined by
\begin{equation}
  \label{eq:eta_torus_volume}
  e^{-\eta} = \int_{T^2}\sqrt{\mathrm{det}(g_{ij})} =
  \sqrt{g_{11}g_{22}}\ .
\end{equation}

As was already mentioned in the previous section, the three-form $\Cd$
does not carry any degrees of freedom in 4 dimensions, and we choose
to integrate it out. The equation of motion for $\Cd$ derived from
\eqref{eq:K3xT2_orienti_action} reads $\rmd (\Cd -g^2\wedge \Cd_2) =
0$. Its solution $\Cd = g^2\wedge \Cd_2$ is then inserted back into
\eqref{eq:K3xT2_orienti_action}.  Similarly, the massless two-form
field $\Cd_2$ can be dualized to a massless scalar field $\gamma$.
Following the well-known dualization prescription (see, for instance,
\cite{Louis:2002ny}), we arrive at the following action for $\gamma$
\begin{equation}
  \label{eq:dual_action_gamma}
  S_\gamma = \tfrac{1}{16} \int e^{2\varphihat+2\rho}g^{11}
  (\rmd \gamma +2B^P\rmd\Cd_{1P})
  \wedge *(\rmd \gamma +2B^Q\rmd\Cd_{1Q})\ ,
\end{equation}
which replaces all terms containing $\Cd_2$ in the action
\eqref{eq:K3xT2_orienti_action} .

In the following two sections, we perform the necessary field
redefinitions which bring the action into the canonical form of $N=2$
supergravity.  Let us start with the vector multiplets.

\subsubsection{Vector multiplets}
\label{sec:vector-multiplets-K3T2}
In order to to display the supergravity basis we need to perform a set
of field redefinitions which decouples the kinetic terms of the vector
multiplet scalars from the hypermultiplet scalars.  As was already
mentioned in section \ref{sec:massless-spectrum_orientifold}, the
scalars in the vector multiplets are the $\neven -1$ metric moduli in
$\xi^{3A}$, the $\Cd_{2A}, B_{12}, \eta$, and $e^{-2\varphihat - \rho}g_{22}$.
In order to exhibit the $N=2$ special geometry, we assemble them into
the following $\neven +1$ complex fields
\begin{equation}
  \begin{aligned}\label{eq:def_complexvars_z}
    z^A &= \Cd_2^{\phantom{2}A}+ie^{-\varphihat -\tfrac12 \rho}\sqrt{2g_{22}}\,\xi^{3A}\ ,\\ 
    s&=
    B_{12} + ie^{-\eta} \ ,
  \end{aligned}
\end{equation} 
where $\eta$ is defined in \eqref{eq:eta_torus_volume}.  We recall
that $\xi^{3A}$ has unit norm and thus carries only $\neven -1$
degrees of freedom.  The required extra degree of freedom turns out to
be the factor $e^{-\varphihat-\tfrac12 \rho}\sqrt{2g_{22}}$.  In terms of these fields
the kinetic terms are indeed block diagonal and read for the vector
multiplet scalars $z^A$  and $s$
\begin{align}  \label{eq:scalaraction_vectormultiplets_specialkaehler}
  S_{\mathrm{vector}}= \int \frac{-1}{(s-\bar{s})^2}\,\rmd s \wedge * \rmd \bar{s} 
  + G_{A\bar{B}}\, \rmd z^A\wedge * \rmd
  \bar{z}^B,
\end{align} where the coupling $G_{A\bar{B}}$ is given by the
expression
\begin{equation}
  \label{eq:G_AB}
  G_{A\bar{B}} =
  -4\frac{(z-\bar{z})_A(z-\bar{z})_B}{((z-\bar{z})^C(z-\bar{z})_C)^2} +
  \frac{2\eta_{AB}}{(z-\bar{z})^C(z-\bar{z})_C}\ .
\end{equation}
The combined metric  defined by
\eqref{eq:scalaraction_vectormultiplets_specialkaehler} is  K\"ahler
with the K\"ahler potential
\begin{equation}
 \label{eq:kahlerpotential_vectorscalars}
 \mathcal{K} = - \ln i(\bar{s}-s)
 -\ln[-\tfrac{1}{8}\eta_{AB}(z-\bar{z})^A(z-\bar{z})^B]\ .
\end{equation}
Note that $\mathcal{K}$ can also be expressed in terms of geometrical
quantities as 
\begin{equation}
  \label{eq:Kaehler_potential_geometrical}
  \mathcal{K}= -\ln e^{-2\varphihat}\int_{\K3 \times T^2} \Im \Omega \wedge \rmd y^2 \wedge * (\Im \Omega \wedge \rmd y^2)\, ,
\end{equation}
where we used \eqref{eq:xi_2forms}.
$\mathcal{K}$ is a K\"ahler potential for the coset space 
\begin{equation}
 \label{eq:specialkahler_vectormultiplets}
{\cal M_{\rm v}} = \frac{\mathrm{SU}(1,1)}{\mathrm{U}(1)}
\times\frac{\mathrm{SO}(2,\neven)}{\mathrm{SO}(2)
  \times \mathrm{SO}(\neven)}\ .
\end{equation}
Consistent with $N=2$ supergravity ${\cal M_{\rm v}}$ is a special
K\"ahler manifold in that $K$ can be written in the form  
$\mathcal{K}=-\ln i[\bar{X}^I {\cal F}_I - X^I \bar{{\cal
    F}}_I ]$ for
\begin{equation}
 \label{eq:prepotential}
{\cal F}_I = \partial_I {\cal F}\ , \qquad {{\cal F}}=-\frac{S\eta_{AB}Z^{A}Z^{B}}{2X^0}\ ,
\end{equation}
and a  choice of special coordinates 
$X^I=(X^0,S,Z^A) = \tfrac12(1,s,z^A), I=0,...,\neven+1$. 

 This prepotential ${\cal F}$ also determines the couplings of the
field strengths of the graviphoton and the $\neven +1$ vector fields.  
We label the vector fields with the same
index $I$ and define
\begin{equation}
 \label{eq:canonicalvecfields}
F^I=\rmd A^I = (\rmd g^2, \rmd
 B_1,\rmd\Cd^A)\ .
\end{equation}
 In a consistent $N=2$ supergravity Lagrangian, they
should couple as \cite{Andrianopoli:1996cm}
\begin{equation}
 \label{eq:canonicalvecaction}
S_\mathrm{F}=\frac{1}{2} \int \
 \mathrm{Re}\mathcal{N}_{IJ}F^I\wedge F^J- \mathrm{Im}
 \mathcal{N}_{IJ}F^I\wedge *F^J,
\end{equation}
 where the matrix $\mathcal{N}$ is expressed in terms of the
prepotential $\mathcal{F}$
\begin{equation}
 \label{eq:couplingmatrixfromprepotential}
\mathcal{N}_{IJ} =
 \bar{\mathcal{F}}_{IJ} + 2i\frac{\Im \mathcal{F}_{IJ} \Im
 \mathcal{F}_{JL} X^KX^L}{\Im \mathcal{F}_{MN}X^MX^N}\ .
\end{equation}
In equation \eqref{eq:couplingmatrix} we display the matrix
$\mathcal{N}_{IJ}$ obtained from the effective action
\eqref{eq:K3xT2_orienti_action}.  It is indeed
consistent with the matrix obtained by inserting $\mathcal{F}$ given
in 
\eqref{eq:prepotential} into \eqref{eq:couplingmatrixfromprepotential}.

\subsubsection{Hypermultiplets}
\label{sec:hypermultiplets-K3T2}
Let us now turn to the geometry of the scalar fields in the
hypermultiplets. The field redefinition \eqref{eq:def_complexvars_z}
decoupled the scalars in the vector multiplet so that the remaining
scalar kinetic 
terms in the effective action \eqref{eq:K3xT2_orienti_action} can be
written as
\begin{equation}
  \label{eq:action_hyperscalars_raw}
  \begin{aligned}
    S_{\mathrm{hyper}} = \int &
    -\tfrac{1}{16}\rmd \Htd{P}{Q}\wedge *\rmd \Htd{Q}{P} \\
    &+\tfrac{1}{4}e^{2\varphihat}g^{11}\left(
      \rmd(e^{-\varphihat}\sqrt{g^{11}})
      \wedge * \rmd (e^{-\varphihat} \sqrt{g_{11}}) 
      +\rmd \A_1 \wedge * \rmd \A_1
    \right)\\
    &+\tfrac{1}{4}e^{2\varphihat+2\rho}g^{11}
    \rmd(e^{-\varphihat-\rho}\sqrt{g^{11}})
    \wedge * \rmd (e^{-\varphihat-\rho} \sqrt{g_{11}})\\
    &+\tfrac{1}{4}e^\rho H_{PQ}\rmd B^P \wedge * \rmd B^Q\\
    & +\tfrac{1}{4}e^{2\varphihat+\rho}g^{11}H^{PQ}
    (\rmd \Cd_{1P}+\A_1\rmd B_P)
    \wedge *( \rmd\Cd_{1Q}+\A_1\rmd B_Q)\\
    &+\tfrac{1}{4}e^{2\varphihat+2\rho}g^{11}
    (\tfrac{1}{2} \rmd \gamma + B^P\rmd\Cd_{1P})
    \wedge *(\tfrac{1}{2}\rmd\gamma + B^Q\rmd\Cd_{1Q}).
  \end{aligned}
\end{equation} 
We will now show that this defines a metric on the quaternionic
manifold
\begin{equation}
  \label{eq:target_space_hyper}
  \mathcal{M_{\rm h}}=\frac{\mathrm{SO}(4,\nodd)}{\mathrm{SO}(4)\times \mathrm{SO}(\nodd)}\ .
\end{equation}
To do so we use the fact that  $\mathcal{M_{\rm h}}$ is in the image
of the c-map \cite{Cecotti:1988qn,Ferrara:1989ik}. More specifically
this implies that  $\mathcal{M_{\rm h}}$ can be viewed as a 
fibration over a special K\"ahler base space which, for the case at hand, is 
 the manifold
\begin{equation}
\label{eq:specialkahler_c-map}
\mathcal{M_{\rm b}}=
\frac{\mathrm{SU}(1,1)}{\mathrm{U}(1)} \times
 \frac{\mathrm{SO}(2,\nodd-2)}{\mathrm{SO}(2)\times \mathrm{SO}(\nodd-2)}\ .
\end{equation}
 Indeed, the first 2 lines of
\eqref{eq:action_hyperscalars_raw} are precisely the metric  of 
$\mathcal{M_{\rm b}}$. This follows from our discussion in section
\ref{sec:massless-spectrum_orientifold} and in particular from 
eq.~\eqref{eq:K3_metric_moduli_space_orientifolded}. There we already
argued that the $\Htd{P}{Q}$ can be viewed as the coordinates of 
the second factor of $\mathcal{M_{\rm b}}$.
Furthermore, we recognize the second line of
\eqref{eq:action_hyperscalars_raw} as a 
standard parametrization of the coset $\mathrm{SU}(1,1)/\mathrm{U}(1)$ by
combining $\A_1$ and $e^{-\varphihat}\sqrt{g_{11}}$ into
the complex field 
\begin{equation}
 \label{eq:S_su11coset_hypermultiplets}
\T := \A_1 + i
 e^{-\varphihat}\sqrt{g_{11}}\ .
\end{equation}
In order to compare the action \eqref{eq:action_hyperscalars_raw}
with the form given in \cite{Ferrara:1989ik}
we define \cite{Andrianopoli:1996cm} 
\begin{equation}
 \label{eq:period_matrix_c-map}
\mathcal{M}_{PQ} = \mathrm{Re} \T\,
 \eta_{PQ} + i \mathrm{Im} \T\, H_{PQ}\ .
\end{equation}
and
\begin{equation}
  \label{eq:phi_phitilde_redefinition_hypermultiplets}
  \phi := 2 e^{-\varphihat-\rho}\sqrt{g_{11}}\ ,
  \qquad \phitilde :=  \gamma + B^P\Cd_{1P}\ .
\end{equation} 
With the help of
\eqref{eq:S_su11coset_hypermultiplets}--\eqref{eq:phi_phitilde_redefinition_hypermultiplets}
we can recast the action \eqref{eq:action_hyperscalars_raw} into the
form
\begin{align}
 \label{eq:action_hyperscalars_c-map}
 S_{\mathrm{hyper}}= 
 \int &\frac{-1}{(\T-\bar{\T})^2}
 \rmd \T \wedge *\rmd \bar{\T} 
 -\tfrac{1}{16} 
 \rmd \Htd{P}{Q} \wedge *\rmd \Htd{P}{Q}\notag \\
 & +\frac{1}{4\phi^2} \rmd\phi\wedge * \rmd \phi 
 + \frac{1}{2\phi}
 (\Im\mathcal{M})_{PQ}\rmd B^P\wedge *\rmd B^Q\notag\\
 & +\frac{1}{2\phi} 
 \left(
   \Im\mathcal{M}
 \right)^{-1\ PQ}
 \left(\rmd
   \Cd_{1P} + (\Re\mathcal{M})_{PR}\rmd B^R
 \right) \wedge * 
 \left(
   \rmd C_{1Q} + (\Re\mathcal{M})_{QS}\rmd B^S
 \right)\notag\\
 & +\frac{1}{4\phi^2}
 (\rmd \phitilde + B^P\rmd \Cd_{1P}-\Cd_{1P}\rmd B^P)
 \wedge *
 (\rmd \phitilde + B^Q\rmd \Cd_{1Q}-\Cd_{1Q}\rmd B^Q)\ ,
\end{align} 
which exactly coincides with the explicit form of the c-map as
given in \cite{Ferrara:1989ik}.

This ends our discussion of type IIA supergravity  compactified on 
orientifolds of $\K3\times T^2$. Our  main result is 
that using a KK-reduction the scalar field space  is
determined to be
 \begin{equation}\label{Mresult2}
{\cal M}\ =\ \frac{\mathrm{SU}(1,1)}{\mathrm{U}(1)}
\times \frac{\mathrm{SO}(2,\neven)}{\mathrm{SO}(2)\times
  \mathrm{SO}(\neven)} \times \frac{\mathrm{SO}(4,\nodd)}{\mathrm{SO}(2)\times
  \mathrm{SO}(\nodd)}\ ,
\end{equation}
where $\neven+\nodd=22$ and 
$\neven (\nodd)$ count the number of even (odd) harmonic
two-forms of K3.


\section{SU(2)-structure orientifolds}
\label{sec:su2-struct-orient}
We are now in a position to discuss the more general case of type IIA
compactification on a generic SU(2) structure manifold.  Before we
move on to the orientifold projection and the effective action, we
recall some facts about six-dimensional SU(2)-structure manifolds and
briefly discuss the moduli space of metrics on these manifolds as
determined in \cite{Triendl:2009ap}.

\subsection{SU(2)-structure manifolds}
\label{sec:su2-struct-manif}

As stated before, the reduction of the structure group of a
six-dimensional manifold $\Y6$ to SU(2) is equivalent to the
existence of two globally defined spinors $\eta^i$ on $\Y6$.  
With the help of 
these spinors, one can define a real 2-form $J$, a complex  2-form
$\Omega$, and a
complex one-form $K$ exactly as in section
\ref{sec:type-iia-orient-k3-times-t2}, i.e.
\begin{equation}
  \label{eq:su2struct_2forms}
  \begin{aligned}
    J &= \tfrac{i}{4}{
      (\eta^{1\dag}_{-}\gamma_{mn}\eta^1_{-} 
      -\eta^{2\dag}_-\gamma_{mn}\eta^2_{-})\,
      \rmd Y^m \wedge \rmd Y^n}\\
    \Omega &= \tfrac{i}{2}\eta^{1\dag}_-\gamma_{mn}\eta^2_-\,
    \rmd Y^m \wedge \rmd Y^n\\
    K &= \eta^{2\dag}_- \gamma_m \eta^1_+\,
    \rmd Y^m = K^2 + i K^1 ,
  \end{aligned}
\end{equation}
where $Y^m, m=1,\ldots,6$ denote the coordinates on $\Y6$.  However,
for a generic $\Y6$ neither of these forms is necessarily closed as
they are for $\K3 \times T^2$.  Using Fierz identities and the
definitions \eqref{eq:su2struct_2forms}, one can show that $J, \Omega$
and $K$ obey \cite{Waldram,BovyLustTsimpis}\footnote{A set of globally
  defined differential forms $K, J$ and $\Omega$ subject to the
  constraints \eqref{eq:su2struct_ortho_constraints},
  \eqref{eq:su2struct_2form_compat_constraints} is an equivalent
  characterization of an SU(2)-structure on a six-dimensional
  manifold.}
\begin{gather}
  \iota_K J =0\, ,\quad
  \iota_K\Omega=\iota_K\bar{\Omega}=0\, , 
  \label{eq:su2struct_ortho_constraints}\\
  \Omega \wedge \bar{\Omega} = 2 J \wedge J \neq 0\, ,\quad \Omega\wedge
  J=0\, ,\quad \Omega \wedge \Omega =
  0\, ,\label{eq:su2struct_2form_compat_constraints}
  \\ 
  K^1_{\phantom{1}m}K^{1m}= 1 = K^2_{\phantom{2}m}K^{2m}\,, \quad K^1_{\phantom{1}m}K^{2m}=0\,\label{eq:su2struct_2dmetric} .
\end{gather}

A generic $\Y6$ with $\sutwo$-structure will no longer be a direct
product $\mathcal{M}_4\times\mathcal{M}_2$, but it follows from the
constraints
\eqref{eq:su2struct_ortho_constraints},~\eqref{eq:su2struct_2form_compat_constraints}
that the tangent bundle still splits into two orthogonal sub-bundles:
the 2-dimensional part $T_2\Y6$ spanned by the components of $K$, and
its orthogonal complement $T_4\Y6$, or in other words, an almost product
structure exists on $\Y6$ \cite{Waldram,Grana:2005sn,BovyLustTsimpis,Triendl:2009ap,Louis:2009dq,Cortes:2003zd}.  As a further consequence the volume form  splits
according to
\begin{equation}
  \label{eq:vol_form_split}
  \begin{aligned}
    \mathrm{vol}_\Y6 = \mathrm{vol}_2 &\otimes \mathrm{vol}_4 \sim
    K\wedge \bar{K} \otimes J \wedge J\ .
  \end{aligned}
\end{equation} 
We will make the extra assumption that the almost-product structure is
integrable, which seems necessary in order to make the calculation of
the effective action tractable \cite{Louis:2009dq}.\footnote{By
  integrability of the almost-product structure, we mean that local
  coordinates $y^i, i=1,2$, $z^a, a=1,...,4$ can be found in every
  neighborhood of $\Y6$, such that $T_2\Y6$ is spanned by the
  $\partial / \partial y^i$, and $T_4\Y6$ is spanned by the
  $\partial/ \partial z^a$. }
  
There is no general procedure by which one can construct a set of
light Kaluza-Klein modes on a general SU(2)-structure background.  On
Calabi-Yau manifolds, there is a clear distinction between the
harmonic modes, which are massless, and the heavier modes, whose
masses are at the Kaluza-Klein scale.  In the non-Calabi Yau case, the
distinction between light and heavy modes in the compactification is
not obvious.  The current procedure is to assume that, nevertheless, a
suitable finite set of ``light'' modes exists, whose properties can
then be constrained by various consistency conditions \cite{GLW,KP}.

Using these assumptions, the scalar field space for the light modes of
SU(2)-structure compactifications was determined in
\cite{ReidEdwardsSpanjaard,Triendl:2009ap,Louis:2009dq}.  It was shown
in \cite{Triendl:2009ap} that in the absence of massive but light
gravitino multiplets, the low-energy theory is determined by a set of
$n$ two-forms $\omega^\alpha,\alpha=1,...,n$ which describe the
deformations of $\Omega$, $J$, and the complex one-form $K=K^2+i K^1$.
These two-forms are the analogue of the 22 harmonic two-forms of
K3. Furthermore, the intersection form $\eta^{\alpha\beta}$ is defined
as in \eqref{eq:intersectionproduct_K3} and can be shown to have
signature $(3,n-3)$ instead of $(3,19)$ for K3.  The deformation space
of $\Omega$ and $J$ is again a symmetric space analogous to the moduli
space of K3 metrics given in \eqref{eq:K3metricmoduli} and was found
to be \cite{Triendl:2009ap}
\begin{equation}
  \label{eq:su2struct_metric_moduli_4d}
\mathcal{M}_{J,\Omega}=
  \frac{\mathrm{SO}(3,n-3)}{\mathrm{SO}(3)\times \mathrm{SO}(n-3)}\times \Reals^+\ .
\end{equation}
The deformations corresponding to the metric on $T_2\Y6$ are parametrized by
its components $\ghat_{ij}$ defined by the line element 
$\hat{ds}^2_{T_2\Y6}= \ghat_{ij}K^iK^j$. They again span the coset
space  
\begin{equation}
  \label{eq:su2struct_metric_2d}
  \mathcal{M}_{K}=
  \frac{\mathrm{SU}(1,1)}{\mathrm{U}(1)}\ .
\end{equation}

As we already said, $K, \Omega$ and $J$ are no longer necessarily
$\rmd$-closed on $\Y6$ and the exterior derivatives parametrize the
(intrinsic) torsion of the manifold. Imposing that the truncation to a
finite set of light modes is non-degenerate,
constrains the structure of the torsion terms. For the case at hand
one has \cite{ReidEdwardsSpanjaard,Louis:2009dq}
\begin{subequations}
  \label{eq:form_exterior_derivsI}
  \begin{align}
    \rmd \omega^\alpha &=
    \Dtd{i}{\alpha}{\beta}K^i\wedge\omega^\beta\ ,\qquad \alpha, \beta = 1, \ldots, n\ ,\\ 
    \rmd K^i &= \theta^i K^1 \wedge K^2\, ,  \qquad i = 1,2\ ,
  \end{align}
\end{subequations}
where $\Dtd{i}{\alpha}{\beta}$ and $\Dtd{i}{j}{k}$ are
constant.\footnote{Exterior derivatives of the form $\rmd K^i =
  D^i_{\phantom{i}\alpha}\omega^\alpha$ can be ruled out as a
  consequence of the integrable almost-product structure.}  Imposing
$\rmd^2 = 0$ and $\int \rmd
(K^i\wedge\omega^\alpha\wedge\omega^\beta)=0$ implies the following
constraints
\begin{subequations}
  \label{eq:constraints_exterior_derivs}
  \begin{align}
    \Dtd{i}{\alpha}{\gamma}\Dtd{j}{\gamma}{\beta}
    -\Dtd{j}{\alpha}{\gamma}\Dtd{i}{\gamma}{\beta} 
    &= \epsilon_{ij}\theta^k\Dtd{k}{\alpha}{\beta}\ ,\\
    \Dtd{i}{\alpha}{\gamma}\eta^{\gamma\beta}+\eta^{\alpha\beta}\epsilon_{ij}\theta^j 
    &= -\eta^{\alpha\gamma}\Dtd{i}{\beta}{\gamma}\ .
  \end{align}
\end{subequations}
The constraint (\ref{eq:constraints_exterior_derivs}b) implies that we
can define traceless $(n\times n)$ matrices $T_{i}$ as
\begin{equation}
  \label{eq:traceless_generators}
  \Ttd{i}{\alpha}{\beta} =
  \Dtd{i}{\alpha}{\beta}+\tfrac12\epsilon_{ij}\theta^j\deltatd{\alpha}{\beta} \, .
\end{equation}
In terms of the $T_i$, the constraints
(\ref{eq:constraints_exterior_derivs}) take the form
\begin{equation}
  \label{eq:constraints_traceless_generators}
  \begin{aligned}[]
    [T_i,T_j] &= \epsilon_{ij}\theta^k T_k\, ,\\
    \Ttd{i}{\alpha}{\gamma}\eta^{\gamma\beta}&= -\eta^{\alpha\gamma}\Ttd{i}{\beta}{\gamma}\, ,
  \end{aligned}
\end{equation}
which implies that the $T_i$ are in the algebra of SO(3,$n$-3).  For
completeness, we also rewrite the exterior derivatives
(\ref{eq:form_exterior_derivsI}) in terms of the $T_i$.
\begin{equation}
  \label{eq:form_exterior_derivs}
  \begin{aligned}
    \rmd \omega^\alpha &=
    \Ttd{i}{\alpha}{\beta}K^i\wedge\omega^\beta 
    +\tfrac12\theta^i\epsilon_{ij}K^j\wedge \omega^\alpha 
    \ ,\qquad \alpha, \beta = 1, \ldots\, ,n,\\
    \rmd K^i &= \theta^i K^1 \wedge K^2\, ,  \qquad i = 1,2\ .
  \end{aligned}
\end{equation}

Let us now implement the orientifold projection on such generic
SU(2) structure backgrounds.

\subsection{Orientifold projection}
\label{sec:orient-proj-su2struct}
We have argued in the previous section that under reasonable
assumptions the space of light Kaluza-Klein modes in SU(2)-structure
compactifications can be constructed, and that this space has a
similar structure as its $\K3 \times T^2$ counterpart.  In particular
the massless modes of $\K3\times T^2$ compactifications are replaced
by a finite set of light modes with similar couplings.

{} Under these assumptions it is straightforward to also generalize the
orientifold projection, which is the topic of this section.  In
particular in both cases ($\K3 \times T^2$ and SU(2)-structure
manifolds) we have given $J, \Omega$ and $K$ in terms of spinor
bilinears in eqs.~\eqref{eq:su2struct_2forms_K3xT2}, \eqref{Kdef} and
\eqref{eq:su2struct_2forms}. Furthermore the orientifold projections
\eqref{eq:sigma_compl_struct_hol_2form} and
\eqref{eq:orientifold_involution_torus} were derived from the action
of the orientifold map $\sigma$ on the two globally defined spinors
$\eta^i$ given in \eqref{eq:involution_spinors}.  Therefore we can
immediately conclude
\begin{equation}
  \begin{aligned}
    \label{eq:sigma_on_forms_su2struct}
    \sigma^*(J) &= -J\ ,\qquad
    \sigma^*(\Omega) &= -\bar{\Omega}\ ,\qquad
    \sigma^*(K) &= \bar{K}\ .
  \end{aligned}
\end{equation}
Correspondingly, the considerations from section
\ref{sec:massless-spectrum_orientifold} still apply.  The generalized
space of Kaluza-Klein modes is divided into $\sigma^*$-even and -odd
modes as before, with signature of the intersection forms on the
$H^{2,+}$ and $H^{2,-}$ equal to $(1,\neven -1)$ and $(2,\nodd-2)$.
For our purposes, the only difference is that the number $n=\nodd
+\neven$ of ``light'' two-forms in the Kaluza-Klein expansion is now
arbitrary, depending on the details of the internal manifold $\Y6$.
Thus, the moduli space of metrics
\eqref{eq:su2struct_metric_moduli_4d} on $T_4\Y6$ is reduced to
\begin{equation}
  \label{eq:su2struct_metric_moduli_orientifolded}
  \mathcal{M} = \frac{\mathrm{SO}(1,\neven-1)}{\mathrm{SO}(\neven-1)} \times
  \frac{\mathrm{SO}(2,\nodd-2)}{\mathrm{SO}(2) \times \mathrm{SO}(\nodd-2)} \times \Reals^+\ ,
\end{equation}
exactly as in \eqref{eq:K3_metric_moduli_space_orientifolded}.  On
$T_2\Y6$ the metric degrees of freedom are again reduced to the
diagonal components $g_{11}, g_{22}$.

To determine the projection of the remaining modes, we can truncate
the Kaluza-Klein expansion exactly as we did in section
\ref{sec:massless-spectrum_orientifold}. Therefore the structure of
the light multiplets and their kinetic terms is completely unchanged.
In particular the scalar field space is still given by
\eqref{Mresult2} (again with $\neven + \nodd$ arbitrary).  The
difference only arises from the non-vanishing torsion components or in
other words from the non-vanishing exterior derivatives given in
\eqref{eq:form_exterior_derivs}.

All that remains to be done, then, is to specify the transformation
properties of the exterior derivatives with
respect to $\sigma^*$.  Since $\sigma^*$ and $\rmd$ commute, a
$p$-form and its exterior derivative must have the same parity.  This
implies that the general form  of the
exterior derivatives given in 
\eqref{eq:form_exterior_derivs} reduces to
\begin{equation}
  \label{eq:even_form_exterior_derivs}
  \begin{aligned}
    \rmd \omega^A &= \Ttd{2}{A}{B}K^2 \wedge \omega^B 
    +\tfrac12\theta K^2\wedge \omega^A
    +\Ttd{1}{A}{Q}K^1 \wedge \omega^Q\ ,\qquad A,B = 1,\ldots,\neven\, ,\\ 
\rmd K^2 &= 0\ ,
  \end{aligned}
\end{equation}
 for the even forms, whereas for the odd forms we have 
\begin{equation}
  \label{eq:odd_form_exterior_derivs}
  \begin{aligned}
    \rmd \omega^P &= \Ttd{2}{P}{Q}K^2 \wedge \omega^Q 
    +\tfrac12\theta K^2\wedge\omega^P
    +\Ttd{1}{P}{B}K^1 \wedge \omega^B\ ,\qquad P,Q = 1,\ldots,\nodd\, ,\\ 
\rmd K^1 &= \theta K^1\wedge K^2\ ,
  \end{aligned}
\end{equation}
where we have omitted the index on $\theta^1$, since $\theta^2=0$.
These exterior derivatives will induce a scalar potential and give
charge to some of the scalar fields. These modifications are the
subject of the next section.

\subsection{Effective action}
\label{sec:effective-action-su2struct}
We are now prepared to discuss the effective action of type IIA
supergravity compactified on a general SU(2)-structure manifold with
orientifold projection.  On a formal level, the only differences with
the compactification on $\K3 \times T^2$ are that the forms $K^i$
replace the differentials $\rmd y^i$ in the Kaluza-Klein expansion
\eqref{eq:KK_expansion_formfields},
\eqref{eq:10D_metric_decomposition}, as well as the fact that the
expansion forms $K^i, \omega^\alpha$ are no longer required to be
closed.  Physically, the effect of choosing a background manifold with
intrinsic torsion is that the fields parametrizing its deformations
become charged.  This leads to an effective action with gauge
symmetries and a corresponding potential for the scalar fields.  In
case of an SU(2)-structure compactification, the effective action is
an $N=4$ gauged supergravity \cite{ReidEdwardsSpanjaard,DMLST}.  If in addition the orientifold
projection discussed in the previous section is implemented this $N=4$
theory is reduced to a gauged $N=2$ supergravity.

We now substitute the Kaluza-Klein expansion
\eqref{eq:KK_expansion_formfields} for the modes which survive the
orientifold projection (and which are recorded in
Table~\ref{tab:K3_orienti_spectrum}) into the type IIA effective
action \eqref{eq:typeIIA_eff_action}.  Using the exterior derivatives
given in eqs. \eqref{eq:even_form_exterior_derivs} and
\eqref{eq:odd_form_exterior_derivs}, we obtain an effective action of
the form
\begin{equation}
  \label{eq:eff_action_su2struct_orienti}
  S = S_{\mathrm{kin}}^{(\rmd
    \rightarrow D)} + S_{\mathrm{pot}} \ ,
\end{equation}
where the first term $S_{\mathrm{kin}}^{(\rmd \rightarrow D)}$
coincides with the action given in
eq. ~\eqref{eq:K3xT2_orienti_action}, but the ordinary derivatives for
the following fields are replaced by the covariant derivatives
\begin{equation}
  \label{eq:covderivs_su2struct_orienti}
  \begin{aligned}
    D e^{-\eta} &= \rmd e^{-\eta} -g^2\theta e^{-\eta}\,,\\
    D g_{11} &= \rmd g_{11} -g^2\theta g_{11}\,,\\
    D e^{-\rho} &= \rmd e^{-\rho} +g^2\theta e^{-\rho}\,,\\
    D \Htd{A}{B} &= \rmd \Htd{A}{B}
    -g^2(\Ttd{2}{A}{C}\Htd{C}{B}-\Htd{A}{C}\Ttd{2}{C}{B})\,,\\
    D \Htd{P}{Q} &= \rmd \Htd{P}{Q}
    -g^2(\Ttd{2}{P}{R}\Htd{R}{Q}-\Htd{P}{R}\Ttd{2}{R}{Q})\,,\\
    D B_{12} &= \rmd B_{12}
    -g^2\theta B_{12} +B_{1}\theta ,\\
    D B_P &= \rmd B_P +g^2\Ttd{2}{Q}{P}B_Q +\tfrac12 \theta B_P\,,\\
    D \A_1 &= \rmd \A_1 -g^2\theta \A_1\,,\\
    D \Cd_{2A} &= \rmd \Cd_{2P} +g^2(\Ttd{2}{C}{A}
    -\tfrac12\theta \deltatd{C}{A})\Cd_{2C}\,,\\
    D \Cd_{1P} &= \rmd \Cd_{1P} +g^2(\Ttd{2}{Q}{P}
    -\tfrac12\theta \deltatd{Q}{P})\Cd_{1Q}-\Cd^{A}\eta_{AB}\Ttd{1}{B}{P}\,.
  \end{aligned}
\end{equation}
Furthermore, the Abelian vector field strengths $F^I$ given in
eq.~\eqref{eq:canonicalvecfields} are replaced by the non-Abelian
field strengths
\begin{equation}
  \begin{aligned}
    \label{eq:vectorfieldstrengths_su2struct}
    F^0 &= \rmd g^2\ ,\\
    F^1 &= \rmd B_1-\theta g^2 \wedge B_1\ ,\\
    F^{A+1} &= \rmd C^A
    -g^2\wedge(\Ttd{2}{A}{B}C^B -\tfrac12\theta C^A)\ .
  \end{aligned}
\end{equation}
The additional term $S_{\mathrm{pot}}$ in
\eqref{eq:eff_action_su2struct_orienti} corresponds to the scalar
potential
\begin{equation}
  \label{eq:eff_action_potential_su2struct_orienti}
  \begin{aligned}
    S_\mathrm{pot}=-\int &\tfrac18{e^{2\varphihat+2\rho+\eta}} 
    \left(
      g^{22}H^{PQ}(\Ttd{2}{R}{P}+\tfrac12\theta\deltatd{R}{P})
      (\Ttd{2}{S}{Q}+\tfrac12\theta\deltatd{S}{Q})
    \right.\\
    &\left.\qquad\qquad\qquad
      +g^{11}H^{AB}\Ttd{1}{R}{A}\Ttd{1}{S}{B}
    \right) B_RB_S\\
    &-\tfrac1{32}{e^{2\varphihat+\rho+\eta}}g^{22} \Big(
    [H,D_2]^P_{\phantom{P}Q}[H,T_2]^Q_{\phantom{Q}P}
    +[H,D_2]^A_{\phantom{A}B}[H,T_2]^B_{\phantom{B}A}
    \Big)\\
    &-\tfrac1{16}{e^{2\varphihat+\rho+\eta}}g^{11}
    (\Ttd{1}{A}{Q}\Htd{Q}{P} -\Htd{A}{B}\Ttd{1}{B}{P})
    (\Ttd{1}{P}{C}\Htd{C}{A} -\Htd{P}{R}\Ttd{1}{R}{A})\\
    &+\tfrac5{16}{e^{2\varphihat+\rho+\eta}}g^{22}\theta^2 
    +\tfrac{1}{8}e^{4\varphihat+3\eta+\rho}\theta^2(\A_1)^2\\
    &+\tfrac1{8}{e^{4\varphihat+3\eta+2\rho}}H^{PQ}
    \Big(
    \Cd_{2A}\Ttd{1}{A}{P}
    +\theta\Cd_{1P} (\Cd_{1R}+\A_1B_R)(\Ttd{2}{R}{P}+\tfrac12\theta\deltatd{R}{P})
    \Big)\\ 
  & \qquad
    \cdot \Big(
    \Cd_{2B}\Ttd{1}{B}{Q}
    +\theta\Cd_{1Q} (\Cd_{1S}+\A_1B_S)(\Ttd{2}{S}{Q}+\tfrac12\theta\deltatd{S}{Q})
    \Big) \\
    -\int& (\rmd \Cd -\rmd g^2\wedge \Cd_2)
    \wedge B^P(\Ttd{1}{B}{P}\Cd_{2B}
    -\Ttd{2}{Q}{P}\Cd_{1Q}+\tfrac12\theta\Cd_{1P}).
  \end{aligned}
\end{equation}
We can rewrite the last line of
\eqref{eq:eff_action_potential_su2struct_orienti} into a more standard
form if we integrate out the three-form $\Cd$ as in section
\ref{sec:effective-action-K3T2}. Again, $\Cd$ has no independent
degrees of freedom, but due to the extra topological term which now
arises in its action, a contribution to the potential remains after
its elimination.  Solving the equations of motion and substituting the
result back into the action, we obtain the new term
\begin{equation}
  \label{eq:dual_potential_threeform}
  -\tfrac1{8}\int {e^{4\varphihat+3\eta+3\rho}}
  \left(
    B^P(\Cd_{1Q}\Ttd{2}{Q}{P} -\tfrac12 \theta \Cd_{1P}
    -\Cd_{2A}\Ttd{1}{A}{P})
  \right)^2\\
\end{equation}

As before, the next step is to rewrite the effective action
\eqref{eq:eff_action_su2struct_orienti} in terms of the canonical
$N=2$ field variables.  Since the kinetic terms are unchanged, we use
exactly the same redefinitions
\eqref{eq:def_complexvars_z}, \eqref{eq:S_su11coset_hypermultiplets},
and \eqref{eq:phi_phitilde_redefinition_hypermultiplets} from
section \ref{sec:effective-action-K3T2}.  The local gauge symmetries
which are implicit in the covariant derivatives given in
\eqref{eq:covderivs_su2struct_orienti} can be related to an
appropriate set of Killing vectors on the scalar manifolds
\eqref{Mresult2}.  Let us start with the vector multiplets.

\subsubsection{Vector multiplets}
\label{sec:vector-multiplets}
Using again the field redefinitions given in
\eqref{eq:def_complexvars_z} the kinetic term for the scalar fields in
the vector multiplets read
\begin{align}
  \label{eq:kinetic_term_specialkaehler_su2struct}
  S_{\mathrm{vector}}^{\dtoD}= 
  \int \frac{-1}{(s-\bar{s})^2}\, D s \wedge * D \bar{s} 
  +G_{A\bar{B}} D z^A\wedge * D \bar{z}^B,
\end{align} where $G_{A\bar{B}}$ is given in \eqref{eq:G_AB}, and the
covariant derivatives read
\begin{equation}
  \begin{aligned}
    \label{eq:cov_devs_vectorscalars}
    D_\mu s &= \partial_\mu s -
    \gtd{2}{\mu} \theta s +B_{1\mu}\theta ,\\
    D_\mu z^A &= \partial_\mu z^A -\gtd{2}{\mu}
    (\Ttd{2}{A}{B}z^B-\tfrac12\theta z^A)
    +\Cd_\mu^{\phantom{\mu}B}\Ttd{2}{A}{B}
     -\tfrac12\Cd_\mu^{\phantom{\mu}A}\theta\ .
  \end{aligned}
\end{equation}
 We can combine these covariant derivatives into the form
\begin{equation}
  \label{eq:Killingvec_covdev}
  D_\mu z^i = \partial_\mu z^i -A^I_\mu k^i_I\ ,
\end{equation}
where $z^i$ denotes collectively all vector multiplet scalars $z^i =
(s, z^A)$ and $A^I_\mu$ denotes all gauge fields, i.e.\
$A^I_\mu=(\gtd{2}{\mu},B_{1\mu},\Cd^A_\mu)$.  Comparing
\eqref{eq:Killingvec_covdev} with \eqref{eq:cov_devs_vectorscalars} we
can read off the Killing vectors $k_I = k^i_I\partial_{z^i}$
\begin{equation}
  \label{eq:killingvectors_specialkaehler}
  \begin{aligned}
    k_0 &= \theta s\partial_s + \Ttd{2}{A}{B}z^B\partial_{z^A}
    -\tfrac12\theta z^A\partial_{z^A}\ ,\\
    k_S &= -\theta \partial_s\ ,\\
    k_{A} &= -\Ttd{2}{B}{A}\partial_{z^B}
    +\tfrac12\theta\partial_{z^A}\ .
  \end{aligned}
\end{equation}
This leads to the gauge algebra
\begin{subequations}
  \label{eq:algebra}
  \begin{align}
    [k_{0} , k_S] &= -\theta k_S\ ,\label{eq:commutator_0S}\\
    [k_0,k_A] &= -(\Ttd{2}{B}{A} -\tfrac12\theta\deltatd{B}{A})\, k_B\ ,
    \label{eq:commutator_0A}\\
    [k_S,k_A] &= [k_A,k_B] =0.\label{eq:commutator_ABS}
  \end{align}
\end{subequations}
This solvable algebra is the semi-direct sum of the Abelian algebra of
the translation generators $k_S,k_A$ and the generator $k_0$.
Furthermore, one can check that the non-Abelian field-strengths given
in \eqref{eq:vectorfieldstrengths_su2struct} are indeed of the form
$F^I = d A^I + f^I_{JK} A^J A^K$ for the structure constants defined
via $[k_J,k_K] = f^I_{JK} k_I$.

For completeness let us also compute the (real) Killing prepotentials
$P_I$ which exist for all isometries of a special K\"ahler manifold.
They are defined by \cite{Andrianopoli:1996cm}
\begin{equation}\label{eq:Killing_prepot_sk}
	k^i_I=ig^{i\bar{\jmath}}\partial_{\bar{\jmath}}P_I\ .
\end{equation}
Integrating \eqref{eq:Killing_prepot_sk} for the Killing vectors
\eqref{eq:killingvectors_specialkaehler} we find
\begin{equation}
  \label{eq:Killing_prepotentials_SK}
  \begin{aligned}
    P_0 &=-\tfrac{i}{2}\,\theta\,\frac{s+\bar{s}}{s-\bar{s}}
    -2i\, \frac{\bar{z}_A\Ttd{2}{A}{B}z^B}{(z-\bar{z})^2} 
    -\tfrac{i}{2}\,\theta\,\frac{z^2-\bar{z}^2}{(z-\bar{z})^2}\ ,\\
    P_S &=-i\,\theta\,\frac{1}{s-\bar{s}}\ ,\\ 
    P_A &= -2i \,\frac{(z-\bar{z})_B\Ttd{2}{B}{A}}{(z-\bar{z})^2}-i \theta\, \frac{(z-\bar{z})_A}{(z-\bar{z})^2}.
  \end{aligned}
\end{equation}

\subsubsection{Hypermultiplets}
\label{sec:hypermultiplets}
The scalars in the hypermultiplets are also charged, as can be seen
from the covariant derivatives given in
\eqref{eq:covderivs_su2struct_orienti}.  Using again the definitions
\eqref{eq:S_su11coset_hypermultiplets}--\eqref{eq:phi_phitilde_redefinition_hypermultiplets}
the kinetic terms of the hypermultiplet scalars are given by
\begin{align}
  \label{eq:kinetic_hypermultiplets_su2struct}
 S^{\dtoD}_{\mathrm{hyper}}= \int &\frac{-1}{(\T-\bar{\T})^2}\, D \T \wedge
 * D \bar{\T} - \frac{1}{16}D \Htd{P}{Q}\wedge * D \Htd{Q}{P}\notag \\
  & + \frac{1}{4\phi^2} \rmd \phi\wedge * \rmd \phi + \frac{1}{2\phi}
  (\Im\mathcal{M})_{PQ}D B^P\wedge * D B^Q\notag\\ 
  & +\frac{1}{2\phi}
  \left(\Im\mathcal{M}\right)^{-1\ PQ}\left(D \Cd_{1P} +
    (\Re\mathcal{M})_{PR}D B^R\right) \wedge * \left(D C_{1Q} +
    (\Re\mathcal{M})_{QS} D B^S\right)\notag\\ 
  & + \frac{1}{4\phi^2}(D
  \phitilde + B^P D \Cd_{1P}-\Cd_{1P} D B^P)\wedge * (D \phitilde +
  B^Q D \Cd_{1Q}-\Cd_{1Q} D B^Q)\ ,
\end{align} 
with the covariant derivatives
\begin{equation}
  \begin{aligned}
    \label{eq:covdevs_hyperscalars}
    D_\mu\xi^{iP} &= \partial_\mu \xi^{iP}
    -\gtd{2}{\mu}\Ttd{2}{P}{Q}\xi^{iQ}\ ,\\
    D_\mu \T &= \partial_\mu \T -\gtd{2}{\mu}\theta\T\ ,\\
    D_\mu B^P &= \partial_\mu B^P 
    -\gtd{2}{\mu}(\Ttd{2}{P}{Q}B^Q-\tfrac12\theta B^P)\ ,\\
    D_\mu \Cd_{1P}&= \partial_\mu \Cd_{1P} +\gtd{2}{\mu}(\Ttd{2}{Q}{P}\Cd_{1Q}
    -\tfrac12\theta\Cd_{1P})-\Cddt{\mu}{A}\eta_{AB}\Ttd{1}{B}{Q}\ ,\\
    D_\mu \phitilde &= \partial_\mu \phitilde
    -\Cd_{\mu}^{\phantom{\mu}A}\eta_{AB}\Ttd{1}{A}{P}B^P\ .
  \end{aligned} 
\end{equation}
These covariant derivatives can again be cast into the generic form
$D_\mu q^u = \partial_\mu q^u -A^I_\mu k^u_I,$ where $q^u$
collectively denote all scalars in the hypermultiplets.  Comparing
with \eqref{eq:covdevs_hyperscalars} determines the Killing vectors
$k_I = k^u_I\partial_u$ on the quaternionic manifolds.  We find that
the non-trivial Killing vectors on $\mathcal{M}_\mathrm{h}$ are
\begin{equation}
  \label{eq:Killing_vecs_quaternionic}
  \begin{aligned}
    k_0 &= \Ttd{2}{P}{Q}\xi^{iQ}\partial_{\xi^{iP}}+(\Ttd{2}{P}{Q}B^Q
   -\tfrac12\theta B^P)\partial_{B^P}\\ 
    &\qquad
    -(\Ttd{2}{Q}{P}\Cd_{1Q}-\tfrac12\theta\Cd_{1P})\partial_{\Cd_{1P}}
    +\theta\T\partial_\T\ ,\\
    k_A &= \eta_{AB}\Ttd{1}{B}{P}(\partial_{\Cd_{1P}} + B^P\partial_{\phitilde})\ .
  \end{aligned}
\end{equation}
Obviously, consistency requires that they form the same gauge algebra
as the algebra \eqref{eq:algebra} of the Killing vectors on the
special K\"ahler manifold.  $k_S$ does not act on the quaternionic
space and therefore the only non-trivial commutator we need to check
is $[k_0,k_A]$.  Using in turn the commutation property from
\eqref{eq:constraints_traceless_generators} and the fact that the
$T_i$ are in the algebra of SO(3,3-$n$), we obtain
\begin{equation}
  \label{eq:commutator_quaternionic}
  \begin{aligned}
    [k_0,k_A] =&\,
    \eta_{AB}\Ttd{1}{B}{P}(\Ttd{2}{P}{Q}\partial_{\Cd_{1P}}
    -\tfrac12\theta\partial_{\Cd_{1P}})\\
    &+(\Ttd{2}{P}{Q}B^Q -\tfrac12\theta B^P)
    \eta_{AB}\Ttd{1}{B}{P}\partial_\phitilde\\
    =&\,\eta_{AB}(\Ttd{2}{B}{C} +\tfrac12\theta\deltatd{B}{C})
    \Ttd{1}{C}{P}(\partial_{\Cd_{1P}} +B^P\partial_{\phitilde})\\
    =& -(\Ttd{2}{B}{A}-\tfrac12\theta\deltatd{B}{A})\,\eta_{BC}\Ttd{1}{C}{P}
    (\partial_{\Cd_{1P}}+B^P\partial_{\phitilde})\, ,
  \end{aligned}
\end{equation}
which is indeed the commutation relation \eqref{eq:commutator_0A}.

On a quaternionic K\"ahler manifold, there is an SU(2) triplet of
Killing prepotentials associated to each isometry.  They are computed
in appendix \ref{sec:calc-kill-prep}.  Finally, checking the agreement
of the potential \eqref{eq:eff_action_potential_su2struct_orienti}
with the corresponding expression of $N=2$ is relegated to appendix
\ref{sec:potential}.  This completes our discussion of the properties
of the effective gauged supergravity.


\section{Conclusions}
\label{sec:conclusions}
In this paper, we have constructed an $O6$ orientifold projection of
type IIA string theory, compactified on a background manifold with
SU(2) structure.  In order to find the correct orientifold projection,
we first studied the simpler case of compactification on $\K3 \times
T^2$, where all moduli remain massless.  Having found the $O6$
orientifold projections that leave intact half of the supersymmetry of
these backgrounds, we found that they could be easily generalized to
projections on generic SU(2) backgrounds.  We then applied these
orientifold projections to the effective $N=4$ theory obtained from
compactifications of type II string theory on SU(2) structure
backgrounds \cite{ReidEdwardsSpanjaard,DMLST}.  We have shown that the
result corresponds to a standard gauged $N=2$ supergravity by
performing the appropriate field redefinitions.  We have seen that in
the supergravity field basis, the multiplets mix the Ramond and
Neveu-Schwarz fields.  The effective theory has a scalar target space
\begin{equation}
  \label{eq:scalartargetspace}
  \frac{\SU(1,1)}{\U(1)} \times 
  \frac{\SO(2,\neven)}{\SO(2)\times\SO(\neven)}
  \times\frac{\SO(4,\nodd)}{\SO(4)\times\SO(\nodd)},
\end{equation}
where $n_\pm$ is the number of 2-forms with even/odd transformations
under the orientifold involution $\sigma$.  Thus, the scalar target
space takes a simple form, but one expects that the last two factors
of \eqref{eq:scalartargetspace} both receive corrections at string
loop order, since they both depend on the dilaton.

Isometries of all sectors of the scalar target space can become
gauged, when the internal manifold has suitable torsion components.
The gauge algebra which we found, is a solvable semi-direct sum of two
Abelian sub-algebras, similar to the algebras found in other
$G$-structure compactifications \cite{Louis:2009dq,Aharony:2008rx}.
These gaugings induce a potential, which is of the canonical form.  An
application would be to investigate moduli stabilization in these
scenarios.

As a next step, we can combine multiple orientifold projections in
order to arrive at a theory with $N=1$ supersymmetry.  If one could
find a further orientifold projection which is still compatible with
some of the gauge transformations, while at the same time it reduces
the supersymmetry, the result would be a simple, yet non-trivial,
$N=1$ toy model.

\subsection*{Acknowledgments}

The work of JL and TD was supported by the Deutsche
Forschungsgemeinschaft (DFG) in the SFB 676 ``Particles, Strings and
the Early Universe''.

We have greatly benefited from conversations with D. Martinez,
B. Spanjaard and H. Triendl.

\clearpage
\appendix


\section{Spinor conventions}
\label{sec:spinor-conventions}
In this appendix we give a brief overview of the conventions used for
the spinor representations in various dimensions, and discuss the
transformation properties of those spinors under the orientifold map.
This section is largely based upon \cite{Koerber:2007hd}, with some
adaptions due to our slightly different conventions.

\subsection{Representations}
\label{sec:spin-repr}
In agreement with the compactification ansatz, the ten-dimensional
spinors transform in a representation of $\Spin (1,3)\times \Spin
(6)$.  The corresponding decomposition of the ten-dimensional
gamma-matrices $\gamma_M$ is given by
\begin{equation}
  \label{eq:gamma_decomposition}
  \Gamma_\mu = \gamma_\mu \otimes \unit\, , 
  \qquad \Gamma_m = \gamma_{5} \otimes \gamma_m\ ,
\end{equation}
where the $\gamma_\mu$ and $\gamma_m$ are the four-dimensional,
respectively six-dimensional gamma-matrices, and $\gamma_{5}$ is the
four-dimensional chirality operator.  The ten-dimensional chirality
operator $\Gamma_{11}$ is the tensor product of the four-dimensional
and six-dimensional chirality operators
\begin{equation}
  \label{eq:10D_chirality_decomposition}
  \Gamma_{11}= \gamma_{5}\otimes\gamma_{7} \, .
\end{equation}
We work with four- and six-dimensional Weyl spinors, and use subscript
$\pm$ to indicate their chirality.  Complex conjugation changes the
chirality, and we have the following Majorana conditions in four and
six dimensions:
\begin{equation}
  \label{eq:4D_6D_majorana}
  \zeta_\pm = B_{(4)}\zeta_\mp^*\, , \qquad \eta_\pm = B_{(6)}\eta_\mp^*\, ,
\end{equation}
where the following relations hold
\begin{subequations}
  \label{eq:4D_6D_conjugation_gammas}
  \begin{align}
    B_{(4)}^{-1}\gamma_\mu B_{(4)} & = \gamma_{\mu}^*\ ,\\
    B_{(6)}^{-1}\gamma_m B_{(6)}&= -\gamma_{m}^*\ .
  \end{align}
\end{subequations}
The ten-dimensional spinors are Majorana-Weyl, and satisfy the Majorana
condition
\begin{equation}
  \label{eq:10D_majorana}
  \varepsilon = B_{(10)}\varepsilon^*\, ,
\end{equation}
where $B_{(10)}$ is given by
\begin{equation}
  B_{(10)} = \Gamma_{11} \cdot B_{(4)}\otimes B_{(6)},
\end{equation}
and satisfies
\begin{equation}
  \label{eq:10D_conjugation_gammas}
  B^{-1}_{(10)} \Gamma_M B_{(10)}= - \Gamma^*_M.
\end{equation}

\subsection{Transformation properties}
\label{sec:spinor-transf-prop}
Locally, the target space involution $\sigma$ is a combination of a
number of reflections.  Since we want to preserve all four-dimensional
symmetry, these reflections will be along directions in the internal
space $\Y6$.  A reflection that preserves the Majorana property and
only acts on the internal component of a ten-dimensional spinor should
act on the spinors with the transformation
\begin{equation}
  \label{eq:reflection_spinors}
  R_m = i \Gamma_m \Gamma_{(10)} = i \unit \otimes \gamma_m\gamma_{(6)}\, .
\end{equation}
For an orientifold with $Op$-planes, $\sigma$ consists of $l=10-(p+1)$
reflections.  Taking the square of $\sigma= R_{m_1}...R_{m_l}$, we get
the following action on spinors
\begin{equation}
  \label{eq:sigma_squared_spinors}
  (\sigma^*)^2 = (-1)^{\frac{l(l-1)}{2}}\unit \, .
\end{equation}
In the case of $O6$ planes, we have $\sigma^2 = -\unit$, which
demonstrates the need for the extra transformation $(-1)^\FL$ in the
orientifold projection $S$.  

If the orientifold projection is to preserve some of the
supersymmetry, $\sigma^*$ must map between the ten-dimensional
supersymmetry parameters
\begin{equation}
  \label{eq:susyrelation}
  \begin{aligned}
    \sigma^*(\varepsilon^\RomI_{10}) & = \varepsilon^\RomII_{10}\, ,\\
    \sigma^*(\varepsilon^\RomII_{10}) & = \pm\, \varepsilon^\RomI_{10}\, ,
  \end{aligned}
\end{equation}
where the minus sign applies in the case of $O6$ orientifolds,
accounting for the fact that $\sigma^2=-1$.

Since $\sigma$ is a symmetry of our chosen background, it must
preserve the spinors $\eta^i$.  Recalling that $(\sigma^*)^2=-\unit$
in the case of an $O6$ orientifold, we are led to the choice
\begin{equation}\label{eq:orienti_internal_spinors}
  \sigma^*(\eta_\pm^i) = \pm\, \eta^i_\mp\, .
\end{equation}
In principle, more general transformations are of course possible, but
in the case of a single orientifold, we can bring the transformation
 into the form
\eqref{eq:orienti_internal_spinors} by a suitable redefinition of the
$\eta^i$.  Looking at the decomposition
\eqref{eq:susy-param-decomposition} of the ten-dimensional
supersymmetry parameters, and using the transformation property
\eqref{eq:orienti_internal_spinors}, we see that dividing out the
relation \eqref{eq:susyrelation} forces 
\begin{equation}
  \varepsilon^\RomI_i = \varepsilon^\RomII_i\, ,
\end{equation}
reducing the available four-dimensional
supersymmetry.

For completeness, we also mention the case of $O4/O8$ orientifolds.
In this case, $(\sigma^*)^2=\unit$, so we do not add $(-1)^\FL$ to the
orientifold action.  With our conventions, we can choose the following
action on the $\eta^i$
\begin{equation}
  \label{eq:orienti_internal_spinors_O4O8}
  \begin{aligned}
    \sigma^*(\eta^1_\pm) &= \pm \eta^2_\mp\, ,\\
    \sigma^*(\eta^2_\pm) &= \mp \eta^1_\mp\, .\\
  \end{aligned}
\end{equation}
In the case of an $O4/O8$ orientifold, the ten-dimensional
supersymmetry parameters are related as in equation
\eqref{eq:susyrelation}, now without the minus sign.  Using
\eqref{eq:orienti_internal_spinors_O4O8} in the decomposition
\eqref{eq:susy-param-decomposition}, we see that the four-dimensional
supersymmetries must satisfy
\begin{equation}
  \begin{aligned}
  \varepsilon^\RomI_1&=\varepsilon^\RomII_2\, ,\\
  \varepsilon^\RomI_2 &= -\varepsilon^\RomII_1\, .
  \end{aligned}
\end{equation}
We see that the presence of an extra internal spinor, i.e.
SU(2) structure, is necessary to define the (supersymmetric)
$O4/O8$ orientifold projection \cite{Koerber:2007hd}. 
Thus this option is absent in the case of orientifolds of 
SU(3)-structure compactifications \cite{GrimmLouis,Benmachiche:2006df}.

\section{Gauge field kinetic couplings}
\label{sec:append-cont-n=2}
Applying equation \eqref{eq:canonicalvecaction} for the canonical form
of the gauge field kinetic term to the effective action
\eqref{eq:K3xT2_orienti_action} obtained from the compactification, we
find that the matrix $\mathcal{N}$ has the following form
\begin{align}
  \mathcal{N}_{IJ} =& \renewcommand{\arraystretch}{1.5}
  \left(\begin{array}{lll}
      -B_{12}\Cd_2^{\phantom{2}A}\Cd_{2A} &
      \frac{1}{2}\Cd_{2}^{\phantom{2}A}\Cd_{2A} & B_{12}\Cd_{2B}\\
      \frac{1}{2}\Cd_{2}^{\phantom{2}A}\Cd_{2A}&0 & -\Cd_{2B}\\
      B_{12}\Cd_{2A}&-\Cd_{2A}& -B_{12}\eta_{AB}
    \end{array} \right)\notag\\ 
  & \renewcommand{\arraystretch}{1.5} +i
  \left(
    \begin{array}{lll}
      -e^{-2\varphihat-\rho-\eta}(g_{22}+g^{11}(B_{12})^2) &
      \multirow{2}{*}{$e^{-2\varphihat-\rho-\eta}g^{11}B_{12}$}
      &\multirow{2}{*}{$e^{-\eta}\Cd_2^{\phantom{2}A}H_{AB}$}\\
      \quad-e^{-\eta}H_{AB}\Cd_{2}^{\phantom{2}A}\Cd_{2}^{\phantom{2}B} &
      &\\ 
      e^{-2\varphihat-\rho-\eta}g^{11}B_{12}& -e^{-2\varphihat-\rho-\eta}g^{11} & 0\\
      e^{-\eta}H_{AB}\Cd_2^{\phantom{2}B}&0 & -e^{-\eta}H_{AB}
    \end{array}
  \right),\label{eq:couplingmatrix}
\renewcommand{\arraystretch}{1}
\end{align} 
which can be written in terms of the complex scalars $s$ and $z^A$
using the field redefinitions \eqref{eq:def_complexvars_z}.  In terms
of the $N=2$ complex variables, the entries of $\mathcal{N}$ become
\begin{equation}
  \label{eq:couplingmatrix_complexfields}
  \begin{aligned}
    \mathcal{N}_{00}&=  -B_{12}\Cd_2^{\phantom{2}A}\Cd_{2A}
    -i\Big(e^{-2\varphihat-\rho-\eta}(g_{22}+g^{11}(B_{12})^2) +
    e^{-\eta}H_{AB}\Cd_{2}^{\phantom{2}A}\Cd_{2}^{\phantom{2}B}\Big)\\
    &=\frac{-1}{2(s-\bar{s})(z-\bar{z})^2}
    \cdot\Big(
    \bar{s}^2(2(\zzbar)^2-2z^2\bar{z}^2) \\
    &\qquad\qquad\qquad+s\bar{s}(4z^2\bar{z}^2-2z^2 (\zzbar) - \bar{z}^2(\zzbar))
    +\tfrac12s^2(z^2-\bar{z}^2)^2\Big)\, ,\\
    \mathcal{N}_{0S}&=\frac{1}{2}\Cd_{2}^{\phantom{2}A}\Cd_{2A} 
    +i e^{-2\varphihat-\rho-\eta}g^{11}B_{12}
    =\frac{1}{4(s-\bar{s})}(s(z^2+\bar{z}^2) -2\bar{s}\zzbar)\, ,\\
    \mathcal{N}_{0A}&= B_{12}\Cd_{2A} + ie^{-\eta}H_{AB}\Cd_2^{\phantom{2}B} 
    =\frac{(s-\bar{s})(z^2-\bar{z}^2)}{2(z-\bar{z})^2}(z-\bar{z})_A
    +\tfrac12\bar{s}(z+\bar{z})_A\, ,\\
    \mathcal{N}_{SS} & = -ie^{-2\varphihat-\rho-\eta}g^{11}
    = -\frac{(z-\bar{z})^2}{4(s-\bar{s})}\, ,\\
    \mathcal{N}_{SA} &=-\Cd_{2A}
    = -\tfrac12(z+\bar{z})_A\, ,\\
    \mathcal{N}_{AB} &= -B_{12}\eta_{AB}-ie^{-\eta}H_{AB}
    =-\bar{s}\eta_{AB}
    -\frac{(s-\bar{s})}{(z-\bar{z})^2}(z-\bar{z})_A(z-\bar{z})_B\, ,
  \end{aligned}
\end{equation}
where we have abbreviated contractions of the $z^A$ and $\bar{z}^A$
with the form $\eta^{AB}$ as $\zzbar, z^2$ and $\bar{z}^2$.  One can
check that the expressions \eqref{eq:couplingmatrix_complexfields}
agree with the result obtained when substituting the prepotential
\eqref{eq:prepotential} into the equation
\eqref{eq:couplingmatrixfromprepotential}.

\section{Calculation of the Killing prepotentials}
\label{sec:calc-kill-prep}
In this appendix, we give some details on the computation of the
Killing prepotentials $P^x_I$ 
on the hypermultiplet target space
$\mathcal{M}_\mathrm{h}=\mathrm{SO}(4,\nodd)/\mathrm{SO}(4)\times
\mathrm{SO}(\nodd)$ following \cite{Andrianopoli:1996cm}.  One can
parametrize the Grassmannian of spacelike 4-planes
${\mathrm{SO}(4,\nodd)/\mathrm{SO}(4)\times \mathrm{SO}(\nodd)}$ using
four orthonormal vectors of dimension $4+\nodd$, which span the
spacelike 4-plane.  Writing these four vectors as the rows of a
$4\times \nodd$-matrix $Z_{au}$, we obtain
\begin{equation}
  \label{eq:cosetmatrix}
  Z^T_{ua}=\frac{1}{\sqrt{2}}\left(
    \begin{array}{lll}
      \begin{array}{l}
        -\frac{1}{2}\sqrt{\frac{\T-\bar{\T}}{\phi}}(\phitilde +
        B^Q\Cd_{1Q})\\
        \quad -\sqrt{\frac{\phi}{\T-\bar{\T}}}\A_1
      \end{array} &\begin{array}{l}
        -\frac{1}{\sqrt{\phi(\T-\bar{\T})}}(\A_1\phitilde\\
        \quad+\Cd_{1Q}\Cd_1^{\phantom{1}Q} +\A_1\Cd_{1Q}B^Q)\\
        \quad +\frac{1}{2}\sqrt{\phi(\T-\bar{\T})}
      \end{array} & -\sqrt{2}\xi^{iQ}\Cd_{1Q}\\ 
      0
      &\frac{2}{\sqrt{\phi(\T-\bar{\T})}} &0\\
      \sqrt{\frac{\phi}{\T-\bar{\T}}}
      -\sqrt{\frac{\T-\bar{\T}}{\phi}}\frac{B^QB_Q}{2}&
      \begin{array}{l}
        \frac{1}{\phi(\T-\bar{\T})}(\phitilde\\
        \quad-B^Q(\Cd_{1Q}+\A_1B_Q))
      \end{array}
      & -\sqrt{2}\xi^{iQ}B_Q\\
      -\sqrt{\frac{\T-\bar{\T}}{\phi}}
      &-\frac{2}{\sqrt{\phi(\T-\bar{\T})}}\A_1 & 0\\
      \sqrt{\frac{\T-\bar{\T}}{\phi}}B_P
      &\frac{2}{\sqrt{\phi(\T-\bar{\T})}}(\Cd_{1P}+\A_1B_P)
      &\sqrt{2}\xi^i_{\phantom{i}P}
    \end{array}\right),
\end{equation} 
where $i=1,2$.  From $Z$, we can compute
the SO(4) component of the connection on $\mathcal{M}_\mathrm{h}$
\begin{equation}
  \label{eq:so4conn_matrix}
  \theta_{ab} = Z_{ua} \eta^{uv} \rmd Z_{vb}\, ,
\end{equation} 
where $\eta$ is the following metric of signature $(4,\nodd)$:
\begin{equation}
  \label{eq:eta_4_20}
\eta_{uv}=\left(
    \begin{array}{ccccc}
      0 & 1 & 0 & 0 & 0\\ 
      1 & 0 & 0 & 0 & 0\\ 
      0 & 0 & 0 &-1 & 0\\ 
      0 & 0 &-1 & 0 & 0\\ 
      0 & 0 & 0 & 0 & \eta_{PQ}
    \end{array}\right).
\end{equation} 
We obtain the SU(2) connection on $\mathcal{M}_\mathrm{h}$ by decomposing
$\theta$ with respect to the three self-dual 't Hooft matrices
$J^{x+}$ given in \cite{Andrianopoli:1996cm}:
\begin{equation}
  \label{eq:so4_connection_decomposition}
  \omega^x= -\tfrac{1}{2}\mathrm{tr}(\theta J^{x+})\, , \qquad
x=1,2,3\ .
\end{equation}
Computing $\theta$ from $Z$ as given in \eqref{eq:cosetmatrix} and
extracting the different components according to
\eqref{eq:so4_connection_decomposition}, we find the SU(2) connection
components $\omega^x$
\begin{equation}
  \label{eq:su2connection}
  \omega^x = \left(
    \begin{array}{l}
      \begin{array}{l}\displaystyle
        \frac{i}{\T-\bar{\T}}\rmd \A_1
        -\frac{1}{2\phi}(\Cd_{1P} \rmd B^P - B^P \rmd \Cd_{1P} 
        -\rmd \phitilde)\\
        \qquad+ \frac{1}{2}(\xi^{1P}\rmd\xi^2_{\phantom{2}P} -
        \xi^{2P}\rmd\xi^1_{\phantom{1}P})
      \end{array} \\ 
      \displaystyle
      \sqrt{\frac{\T-\bar{\T}}{2i\phi}}\xi^1_{\phantom{1}P}\rmd B^P
      -\sqrt{\frac{2i}{\phi(\T-\bar{\T})}}(\A_1
      \xi^2_{\phantom{2}P}\rmd B^P + \xi^{2P}\rmd \Cd_{1P})\\
      \displaystyle
      -\sqrt{\frac{\T-\bar{\T}}{2i\phi}}\xi^2_{\phantom{2}P}\rmd B^P
      -\sqrt{\frac{2i}{\phi(\T-\bar{\T})}}(\A_1
      \xi^1_{\phantom{1}P}\rmd B^P + \xi^{1P}\rmd \Cd_{2P})
    \end{array}\right).
\end{equation} 
The Killing prepotentials $P^x_{I}$, then, are the
solutions to the set of differential equations
\begin{equation}
  \label{eq:diff_eq_killing_prepot}
  -k_I \hook ( \rmd \omega^x 
  +\frac{1}{2} \epsilon^{xyz}\omega^y\wedge\omega^z) 
  = \rmd P^x_{\phantom{x}I} + \epsilon^{xyz} \omega^y
  P^z_{\phantom{z}I},
\end{equation}
where the left-hand side is the insertion of the $I$-th Killing vector
into the SU(2) curvature form, and the right-hand side is the
SU(2)-covariant derivative acting on the triplet $P^x_{I}$.
Combining the connection from equation \eqref{eq:su2connection} with
the Killing vectors $(k_0,k_A)$ \eqref{eq:Killing_vecs_quaternionic}
on the quaternionic manifold, we obtain the following Killing
prepotentials:
\begin{align}
  \label{eq:killing_prepotentials1}
  &\begin{aligned}
    P^1_{0} &=
    \frac{i}{\T-\bar{\T}}\A_1\theta
    -\frac{1}{\phi}\Cd_{1P}(\Ttd{2}{P}{Q}
    -\tfrac12\theta\delta^P_{\phantom{P}Q})B^Q
    +\xi^1_{\phantom{1}P}\Ttd{2}{P}{Q}\xi^{2Q}\, ,\\
    P^2_0 &= 
    \sqrt{\tfrac{\T-\bar{\T}}{2i\phi}}\xi^1_{\phantom{1}P}
    (\Ttd{2}{P}{Q}-\tfrac12\theta\delta^P_{\phantom{P}Q})B^Q\\
    &\qquad-\sqrt{\tfrac{2i}{\phi(\T-\bar{\T})}}
    (\A_1\xi^2_{\phantom{2}P}(\Ttd{2}{P}{Q}
    -\tfrac12\theta\delta^P_{\phantom{P}Q})B^Q
    -\Cd_{1P}(\Ttd{2}{P}{Q}-\tfrac12\theta\delta^P_{\phantom{P}Q})\xi^{2Q})\, ,\\
    P^3_0 &=
    -\sqrt{\tfrac{\T-\bar{\T}}{2i\phi}}\xi^2_{\phantom{2}P}
    (\Ttd{2}{P}{Q}-\tfrac12\theta\delta^P_{\phantom{P}Q})B^Q\\
    &\qquad-\sqrt{\tfrac{2i}{\phi(\T-\bar{\T})}}
    (\A_1\xi^1_{\phantom{1}P}(\Ttd{2}{P}{Q}
    -\tfrac12\theta\delta^P_{\phantom{P}Q})B^Q
    -\Cd_{1P}(\Ttd{2}{P}{Q}-\tfrac12\theta\delta^P_{\phantom{P}Q})\xi^{1Q})\, ,
  \end{aligned}\\
  &\label{eq:killing_prepotentials2}
  \begin{aligned}
    P^1_{A} &=  \eta_{AB}\frac{1}{\phi}\Ttd{1}{B}{P}B^P\, ,\\
    P^2_A &= -\eta_{AB}\sqrt{\frac{2i}{\phi(\T-\bar{\T})}}\Ttd{1}{B}{P}\xi^{2P}\, ,\\
    P^3_A &= -\eta_{AB}\sqrt{\frac{2i}{\phi(\T-\bar{\T})}}\Ttd{1}{B}{P}\xi^{1P}\, .
  \end{aligned}
\end{align}

We can also express these potentials by the following integrals over
the internal manifold:
\begin{align}
  &\begin{aligned}
    P^1_0=&
    -\frac{1}{2}e^{\varphihat+\rho}\sqrt{g^{11}}
    \int_{\Y6}\frac{1}{2}(\rmd(\Re \Omega) \wedge \Re \Omega 
    +\rmd J \wedge J )\wedge \A + \rmd B \wedge \Cd_- \\
    &\qquad+\frac{1}{4}e^{\rho}
    \int_{\Y6}(J \wedge \rmd(\Re \Omega) -\Re \Omega \wedge \rmd J )
    \wedge K^1\, ,\\
    P^2_0=&- \frac{1}{2}e^{\varphihat+\rho}\sqrt{g^{11}}
    \int_{\Y6} \rmd B\wedge \Re \Omega \wedge \A 
    +\Cd_- \wedge \rmd(\Re \Omega)\\
    &\qquad +\frac{1}{2}e^{\frac{3}{2}\rho}
    \int_{\Y6}\rmd B \wedge J \wedge K^1\, ,\\
    P^3_0=& -\frac{1}{2}e^{\varphihat+\rho}\sqrt{g^{11}}
    \int_{\Y6} \rmd B\wedge J \wedge \A 
    -\Cd_- \wedge \rmd J\\
    &\qquad -\frac{1}{2}e^{\frac{3}{2}\rho}
    \int_{\Y6}\rmd B \wedge \Re \Omega \wedge K^1\, ,\\
  \end{aligned}\\
  &\begin{aligned}
    P^1_A&= -\eta_{AB}\frac{1}{2}e^{\varphihat+\rho}\sqrt{g^{11}}
    \int_{\Y6}\rmd B \wedge \omega^B \wedge K^2\, ,\\
    P^2_A &= \eta_{AB}\frac{1}{2}e^{\varphihat+\rho}\sqrt{g^{11}}
    \int_{\Y6}\rmd \Re \Omega \wedge \omega^B \wedge K^2\, ,\\
    P^3_A &= \eta_{AB}\frac{1}{2}e^{\varphihat+\rho}\sqrt{g^{11}}
    \int_{\Y6}\rmd J \wedge \omega^B \wedge K^2\, ,\\
    \end{aligned}
\end{align}
where $\Cd_-$ represents those modes of $\hat{\Cd}$ which are
odd under the action of $\sigma^*$, e.g. ${C_- = \Cd_{1P}K^1\wedge
  \omega^P}$.

\section{The potential}
\label{sec:potential}
In this appendix we check consistency of the potential obtained from
the SU(2)-structure compactification with $N=2$ supergravity. The
latter requires that the potential takes the special form \cite{Andrianopoli:1996cm}:
\begin{equation}
  \label{eq:N=2_SuGra_potential}
  \mathcal{V}=e^{\mathcal{K}}X^I\bar{X}^J
  (g_{\bar{\imath}j}k^{\bar{\imath}}_Ik^j_J+4h_{uv}k^u_Ik^v_J)
  -\left(
    \frac{1}{2}(\Im
    \mathcal{N})^{-1\ IJ}
    +4e^{\mathcal{K}}X^I\bar{X}^J
  \right)
  P^x_IP^x_J,
\end{equation} 
where $g_{\bar{\imath}j}$ and $h_{uv}$ represent the metrics on the
special K\"ahler, resp.\ quaternion-K\"ahler target spaces, the $k_I$
are the Killing vectors from equation
\eqref{eq:killingvectors_specialkaehler}, the $X^I$ are the
homogeneous coordinates on the special K\"ahler manifold and the
$P^x_I$ are the Killing prepotentials associated with the gauged
isometries of the quaternionic manifold.  We now verify that the
potential obtained by Kaluza Klein reduction has this canonical
form.

We start by simplifying the rightmost term in equation
\eqref{eq:N=2_SuGra_potential}.  Inverting the imaginary part of the
$\mathcal{N}$ in \eqref{eq:couplingmatrix} gives
\begin{equation}
  \label{eq:im_N_inverse}
(\Im \mathcal{N})^{-1}=
  e^{2\varphihat+\rho+\eta}\left(
    \begin{array}{lll}
      -g^{22} & -g^{22}B_{12} & -\Cd_2^{\phantom{2}B}g^{22}\\
      -g^{22}B_{12} & -g^{22}(B_{12})^2-g_{11}
      &-g^{22}B_{12}\Cd_2^{\phantom{2}B}\\ 
      -\Cd_2^{\phantom{2}A}g^{22}
      & -g^{22}B_{12}\Cd_2^{\phantom{2}A} &-e^{-2\varphihat-2\rho}H^{AB} 
      -g^{22}\Cd_2^{\phantom{2}A}\Cd_2^{\phantom{2}B}
    \end{array} \right).
\end{equation} 
It follows from the formulas for $\mathcal{K}$ and the definition of
the variables $s$ and $z^A$ in section
\ref{sec:vector-multiplets-K3T2}, that
\begin{equation}
  \label{eq:kaehlerpotential_KKfields}
  \tfrac{1}{2}e^{2\varphihat+\rho+\eta}g^{22}= e^\mathcal{K}.
\end{equation} 
Using equations \ref{eq:im_N_inverse},
\ref{eq:kaehlerpotential_KKfields} and the definition of the
coordinates $X^I$ in section \ref{sec:vector-multiplets-K3T2}, we see
that the contribution from the Killing prepotentials
$P^x_{\phantom{x}I}$ given in
\eqref{eq:killing_prepotentials1},~\eqref{eq:killing_prepotentials2}
reduces to
\begin{align}
  \left(
    \tfrac{1}{2}(\Im \mathcal{N})^{-1\ IJ} +4e^\mathcal{K}X^I\bar{X}^J
  \right) P^x_IP^x_J =&
  \tfrac{1}{2}\eta^{AB}P^x_{\phantom{x}A}P^x_{\phantom{x}B} \ ,
\end{align}
where 
\begin{align}
  \tfrac{1}{2}\eta^{AB}P^x_{\phantom{x}A}P^x_{\phantom{x}B}  
  =\ &\frac{e^\eta}{2\phi^2}\eta_{AB}\Ttd{1}{A}{P}B^P\Ttd{1}{B}{Q}B^Q
  + i\frac{e^\eta}{\phi(\T-\bar{\T})}\eta_{AB}
  (\Ttd{1}{A}{P}\xi^{iP}\Ttd{1}{B}{Q}\xi^{iQ})\notag\\
  =&\tfrac18{e^{2\varphihat+2\rho+\eta}}g^{11}\eta_{AB}
  \Ttd{1}{A}{P}B^P\Ttd{1}{B}{Q}B^Q\notag\\
  &\qquad +\tfrac14{e^{2\varphihat+2\rho+\eta}}g^{11}
  \eta_{AB}(\Ttd{1}{A}{P}\xi^{iP}\Ttd{1}{B}{Q}\xi^{iQ})\, .
  \label{eq:potential_Killing_Prepotentials}
\end{align} 
In the last equation we rewrote the result in terms of the original
variables for ease of comparison with the potential
\eqref{eq:eff_action_potential_su2struct_orienti}. 

Calculating the contribution from the first term in
\eqref{eq:N=2_SuGra_potential} is straightforward.  We insert the
Killing vectors as given in equations
\eqref{eq:killingvectors_specialkaehler}, \eqref{eq:Killing_vecs_quaternionic},
and find
\begin{align}
  \label{eq:potential_Killing_SK}
  e^{\mathcal{K}}X^I\bar{X}^Jg_{\bar{\imath}j}k^{\bar{\imath}}_Ik^j_J
  =& \tfrac{3}{16} e^{2\varphihat+\rho+\eta}g^{22}\theta^2 
  +\tfrac14 e^{2\varphihat+\rho+\eta}g^{22}
  \left(\xi^3_{\phantom{1}A}\Ttd{2}{A}{B}\Ttd{2}{B}{C}\xi^{3C}\right)\, ,
\end{align}
together with
\begin{equation}
  \label{eq:potential_Killing_QK}
  \begin{aligned}
    4e^{\mathcal{K}}X^I\bar{X}^Jh_{uv}k^u_Ik^v_J =
    &\phantom{-}\tfrac18 e^{4\varphihat+3\eta+\rho}\theta^2(\A_1)^2
    +\tfrac18 e^{2\varphihat+\rho+\eta} g^{22}\theta^2\\
    &-\tfrac{1}{32}e^{2\varphihat+\rho+\eta}
    [H,T]^P_{\phantom{P}Q}[H,T]^Q_{\phantom{Q}P}\\
    &+\tfrac{1}{8}e^{2\varphihat+2\rho+\eta}g^{22}
    H^{PQ}B_R(\Ttd{2}{R}{P}+\tfrac12\theta\deltatd{R}{P})
    B_S(\Ttd{2}{S}{Q}+\tfrac12\theta\deltatd{S}{Q})\\
    &+\tfrac18 e^{4\varphihat+2\rho+3\eta}H^{PQ}\\
    &\qquad \cdot\left(
      \Cd_{2A}\Ttd{1}{A}{P}
      -\Cd_{1R}(\Ttd{2}{R}{P} -\tfrac12\theta\deltatd{R}{P})
      -\A_1B_R(\Ttd{2}{R}{P} +\tfrac12\theta\deltatd{R}{P})
    \right)\\ 
    &\qquad \cdot\left(
      \Cd_{2B}\Ttd{1}{B}{Q}
      -\Cd_{1S}(\Ttd{2}{S}{Q} -\tfrac12\theta\deltatd{S}{Q})
      -\A_1B_S(\Ttd{2}{S}{Q} +\tfrac12\theta\deltatd{S}{Q})
    \right)\\
    &+\tfrac14 e^{2\varphihat+\rho+\eta}
    H^{PQ}\xi^3_{\phantom{1}A}\Dtd{1}{A}{P}\xi^3_{\phantom{1}B}\Dtd{1}{B}{Q}\\
    &+\tfrac18 e^{4\varphihat+3\rho+3\eta} \left(
      B^P(\Cd_{2A}\Ttd{1}{A}{P}
      -\Cd_{1R}\Ttd{2}{R}{P}+\tfrac12\theta\Cd_{1P})
    \right)^2\\
    &+\tfrac14 e^{2\varphihat+2\rho+\eta} 
    g^{11}(\xi^3_{\phantom{1}A}\Ttd{1}{A}{P}B^P)^2\, .
  \end{aligned}
\end{equation}
The total potential is now equal to the sum of the contributions
\eqref{eq:potential_Killing_Prepotentials},
\eqref{eq:potential_Killing_SK} and \eqref{eq:potential_Killing_QK},
and most of these terms can be recognized immediately in the potential
\eqref{eq:eff_action_potential_su2struct_orienti} obtained from the
compactification.  The equivalence of the remaining terms can be shown
by rewriting the $\Htd{\alpha}{\beta}$ in terms of the $\xi^{x\alpha}$
using \eqref{eq:reduced_moduli_matrix_xi}, and using the constraints
\eqref{eq:constraints_traceless_generators} on the parameters
$\Ttd{i}{\alpha}{\beta}$.

\providecommand{\href}[2]{#2}\begingroup\raggedright\endgroup

\end{document}